%% file: main.tex
\documentclass[11pt,a4paper]{article}
\usepackage{jcappub} 
\usepackage{amsmath,amssymb}
\usepackage{graphicx}
\usepackage{physics,subcaption}
\usepackage{hyperref,enumitem}         
\usepackage{tabularx} 
\usepackage{pdflscape} 
\usepackage{booktabs}           
\usepackage{array}              

\usepackage{geometry}  
\usepackage{tikz}
\usetikzlibrary{shapes.geometric, arrows.meta, positioning}

\tikzstyle{eqbox} = [rectangle, rounded corners, minimum width=6cm, minimum height=1cm,
draw=blue!70!black, very thick, fill=blue!5, text centered, font=\normalsize]
\tikzstyle{modelbox} = [rectangle, minimum width=4cm, minimum height=0.8cm,
draw=red!70!black, very thick, fill=red!5, text centered, font=\normalsize]
\tikzstyle{arrow} = [thick,->,>=stealth]

\title{Spectral Chebyshev Approximation of Cosmic Expansion in $f(R)$ Gravity}

\author[a]{Akshay Rana}
\affiliation[a]{Department of Physics, St. Stephen's College, University of Delhi, Delhi}
\emailAdd{akshay@ststephens.edu}

\abstract{We present a numerical framework to study the cosmological background evolution in  $f(R)$ gravity by employing a \textit{spectral Chebyshev collocation approach}. Unlike standard integration methods such as Runge--Kutta that often encounter stiffness and accuracy issues, this formulation expands the normalized Hubble function $E(z) = H(z)/H_0$ as a finite Chebyshev series. The modified Friedmann equation is then enforced at selected Chebyshev--Gauss--Lobatto points, converting the original nonlinear differential equation into a system of algebraic relations for the series coefficients. This transformation yields exponentially convergent and numerically stable solutions over the entire redshift domain, $0<z<z_{max}$, eliminating the need for adaptive step-size control. We apply the method to two widely studied $f(R)$ models, Hu--Sawicki and Starobinsky, and perform a combined analysis using cosmic chronometer $H(z)$ data and the Union~3.0 supernova compilation. The reconstructed expansion histories match observations to within $2\sigma$ over $0 < z < 2$, producing best-fit parameters of approximately $(\Omega_{m0}, H_0, \Lambda_{\mathrm{eff}}) \simeq (0.29, 68, 1.2\text{--}2.5\,H_0^2)$. These results indicates that both models reproduce the observed late-time acceleration while permitting small geometric corrections to $\Lambda$CDM. Overall, the spectral Chebyshev method provides a precise and computationally efficient framework for probing modified-gravity cosmologies in the precision-data era.}

\begin{document}
\maketitle
\flushbottom

\section{Introduction}

The discovery that the cosmic expansion is accelerating, first inferred from observations of Type~Ia supernovae~\cite{Riess_1998,Perlmutter_1998}, marked a major turning point in modern cosmology. Subsequent measurements of Baryon Acoustic Oscillations (BAO) ~\cite{Eisenstein_2005,Alam_2016}  and temperature anisotropies in the Cosmic Microwave Background (CMB)~\cite{Spergel_2003,Planck_2018} independently confirmed this accelerated behavior. Within the standard $\Lambda$CDM model, the phenomenon is attributed to a cosmological constant $\Lambda$, interpreted as the energy density of the vacuum. Although this model accurately describes a wide range of observations, it leaves two well-known theoretical puzzles unresolved, namely the cosmological-constant problem and coincidence problem  ~\cite{Peebles_1994,Weinberg_1988, Perivolaropoulos_2022}. These puzzles have motivated the exploration of frameworks in which the late-time acceleration is produced not by an exotic energy component but by a modification of the laws of gravity themselves ~\cite{Clifton_2012,joyce_2015,koyama_2016}.

Among the many possibilities, one of the simplest and most extensively studied is the class of $f(R)$ gravity models. In this approach, the Einstein- Hilbert action is generalized so that the Ricci scalar $R$ is replaced by a general function $f(R)$ ~\cite{Starobinsky_1980,Sotiriou_2010,Felice_2010,romero_2018}. This modification introduces higher-order curvature terms and effectively adds a new scalar degree of freedom, commonly called the \textit{scalaron}, that encapsulates deviations from general relativity. The behavior of this scalar field, determined by the derivatives $f_R = df/dR$ and $f_{RR} = d^2f/dR^2$, controls how the expansion history diverges from the $\Lambda$CDM prediction. At early times, when spacetime curvature is large and matter dominates, viable $f(R)$ models reproduce the standard cosmological evolution. At late times, however, the same framework can lead to accelerated expansion without inserting a cosmological constant explicitly~\cite{Appleby_2007,Bertschinger_2008,Fay_2007}.

The main challange with $f(R)$ gravity lies in solving its cosmological background equations. In general relativity with a cosmological constant, the Friedmann equation is algebraic in the normalized Hubble parameter $E(z)=H(z)/H_0$, so the expansion history can be written down in closed form. By contrast, in $f(R)$ gravity the dynamics are governed by a modified Friedmann equation that involves not only $E(z)$ itself but also its first and second derivatives through the Ricci scalar $R(z)$ and its redshift evolution. This turns the problem into a stiff, nonlinear, second–order differential equation in $E(z)$. As a result, direct numerical integration is often unstable or fails to converge, especially near the $\Lambda$CDM limit where the modifications are small and the equations become very stiff \cite{Hu_2007,Tsujikawa_2008}. In this regime even tiny numerical errors can push the solution away from the physical branch, making it difficult to scan the parameter space reliably. Although these problems are most pronounced close to $\Lambda$CDM, instabilities can also appear more generally because of the nonlinear and second-order nature of the equation \cite{Carloni_2005,Amendola_2007,Felice_2010,Hiroaki_2008}.

A common numerical strategy for solving the background evolution in $f(R)$ gravity  is the direct integration of the modified Friedmann equations using standard ordinary  differential equation (ODE) solvers, such as the Runge--Kutta or Bulirsch--Stoer algorithms \cite{Hu_2007,Amendola_2007,Amendola_2007_2,Hairer_Wanner_1996_book}. Although these methods can, in principle, recover the exact solution, they face severe numerical stiffness due to the highly nonlinear dependence of the Ricci scalar and its derivatives. Moreover, these equations require well-defined initial or boundary conditions - such as $E'(z=0)$ or $R(z=0)$ which are not directly available from cosmological observations and must be assumed or tuned artificially~\cite{Odintsov_2026,Nunes_2017}. As a result, even high-order integrators often fail to converge, demand extremely small step sizes, or drift toward unphysical branches of the solution. The problem becomes particularly acute near the $\Lambda$CDM limit, where the corrections to general relativity are small and numerical noise can dominate the dynamics, leading to unstable or inaccurate results~\cite{Dolgov_2003,Basilakos_2013, Plaza_2025}.

An alternative approach is to employ perturbative methods that expand the solution around the $\Lambda$CDM background. One widely adopted example is the $b$-series technique proposed by Basilakos \textit{et al.}~\cite{Basilakos_2013,Nunes_2017,Plaza_2025}, which expresses the normalized Hubble function $E(z)$ as a power series in the deviation parameter. In the case of the Hu–Sawicki model, the expansion is made in powers of the dimensionless parameter $b$, which quantifies the degree of deviation from General Relativity or the $\Lambda$CDM limit ($b \to 0$ recovers $f(R)=R-2\Lambda$). For the Starobinsky model, the series is instead constructed in powers of $1/R_c$, where $R_c$ represents the characteristic curvature scale that determines the transition between the high- and low-curvature regimes of the theory. Physically, while $b$ measures \textit{how much} the model departs from $\Lambda$CDM, $R_c$ specifies \textit{where} in curvature space these deviations become significant. When these quantities are sufficiently small, the first few terms of the expansion accurately reproduce the background evolution with remarkable numerical efficiency and precision~\cite{Basilakos_2013,xu_2018,Odintsov_2020}. However, the perturbative scheme is intrinsically limited: as the modification parameters increase, higher-order corrections become non-negligible, and the truncated expansion fails to reproduce the true dynamics. This shortcoming, discussed in several follow-up analyses \cite{Felice_2010,Carloni_2015,Odintsov_2021}, leads to systematic underestimation of deviations precisely in the parameter region where $f(R)$ effects are expected to be observationally significant. In short, perturbative expansions are highly efficient for models nearly indistinguishable from $\Lambda$CDM, but their validity quickly deteriorates once nonlinear curvature effects become important.

Another approach available in the literature for  solving the modified Friedmann equation of $f(R)$ cosmologies is offered by \textit{reconstruction-based techniques} \cite{Capozziello_2005,Sultana_2022, Darshan_2025}. Instead of integrating the stiff background equations directly, these approaches start from the observational side: the expansion history $E(z)$ is first reconstructed from data such as Type Ia  supernova luminosity distances, Baryon Acoustic Oscillations (BAO), or cosmic chronometer $H(z)$ measurements. The derived $E(z)$ curve is then examined to determine whether it can be reproduced by a viable $f(R)$ function. In this way, the method avoids the numerical stiffness of the modified Friedmann system and offers a phenomenological bridge between data and theory \cite{Capozziello_2005,Nojiri_2006,Bamba_2012}. However, since the reconstruction procedure does not solve the field equations explicitly, the connection to the underlying model parameters and scalaron dynamics remains indirect. As several studies have pointed out \cite{Nunes_2017,Sultana_2022,Nojiri_2017}, reconstruction methods are best viewed as complementary diagnostic tools—useful for assessing the overall viability of $f(R)$ gravity models against observational data, but not as substitutes for full dynamical integration when accurate parameter estimation is required.

Together, these approaches highlight both the importance and the difficulty of obtaining reliable solutions of the $f(R)$ background equations. Direct integration struggles with stiffness, perturbative expansions fail when deviations become large, and reconstruction methods obscure the underlying model parameters. This combination of challenges motivates the search for alternative techniques that are both mathematically stable and computationally efficient.  

In this work we present such an alternative, based on spectral methods, and in particular on \textbf{Chebyshev polynomial collocation method} \cite{Boyd:2001Spectral,Trefethen:2000Spectral,Canuto:2006Spectral,Shen:2011Spectral}. The central idea is to approximate the normalized Hubble function $E(z)$ by a truncated Chebyshev series defined over the full redshift interval of interest, rather than integrating the background equations step by step as in standard ODE solvers. In the collocation approach, the modified Friedmann equation is not required to hold continuously everywhere, but is instead enforced \emph{exactly at a carefully chosen set of collocation points}. For Chebyshev methods, these are the Chebyshev--Gauss--Lobatto nodes, which cluster near the endpoints of the interval and are known to suppress numerical instabilities such as the Runge phenomenon. By enforcing the equation at these nodes, the original stiff, nonlinear differential equation is transformed into a system of coupled nonlinear algebraic equations for the expansion coefficients of the Chebyshev series. In practice, this means that for any chosen values of model parameters [e.g. $(\Omega_{m0},b, \Lambda)$ in the  Hu–Sawicki Model or $ (\Omega_{m0},R_c,\Lambda$) in the Starobinsky Model], the problem reduces to solving for the finite set of Chebyshev coefficients $\{a_k\}$ that make the background equations hold at all collocation nodes. This nonlinear system is then solved using standard root-finding algorithms such as Newton- Raphson, Powell- hybr, or Newton- Krylov methods \cite{Press:2007NumRecipes,More:1980Hybrid,Kelley:1995NewtonKrylov,Singh_2025}. Each solver offers a different balance between efficiency and robustness, and we comment on their relative performance in this work.

The novelty of this work lies in three key aspects:  
\begin{enumerate}
    \item It introduces spectral Chebyshev collocation as a powerful tool for solving the stiff background equations of $f(R)$ cosmology, offering a global and stable alternative to conventional ODE solvers that often struggle in critical regions of parameter space.  
    \item By reducing the problem to a finite set of Chebyshev coefficients, the method allows the expansion history $E(z)$ to be reconstructed with high accuracy and then evaluated at any redshift in a single step, providing a substantial computational advantage for Bayesian parameter estimation and large-scale MCMC analyses.  
    \item We implement this $f(R)$ solver within an MCMC framework and test it on two benchmark $f(R)$ models; Hu-Sawicki and Starobinsky, which are tested against both $H(z)$ cosmic chronometer data and the Union~3.0 supernova sample. This demonstrates not only the robustness of the approach but also its potential for broader applications in precision cosmology, from structure formation to the study of other modified gravity theories.  
\end{enumerate}

The rest of this paper is organized as follows. In Section~\ref{sec:theory}, we review the basic framework of $f(R)$ gravity, derive the governing background equations in redshift space, and introduce the Hu--Sawicki and Starobinsky models. Section~\ref{sec:Numerical_3} presents the spectral Chebyshev collocation method in detail, including the construction of the Chebyshev expansion, the definition of collocation nodes, and the numerical solution strategy. We then apply this solver to the benchmark models to compute the background expansion history for given sets of model parameters. Section~\ref{sec:data_4} describes the observational datasets used in this analysis, namely the cosmic chronometer $H(z)$ measurements and the Union~3.0 supernova sample. Section~\ref{sec:result_5} integrates the numerical solver into a Bayesian parameter estimation framework and presents the joint constraints obtained from these datasets. Finally, Section~\ref{sec:discussion_5} discusses the implications of our results and outlines possible extensions of the proposed method.

\section{Theoretical Framework}
\label{sec:theory}

In this section we briefly review the theoretical setup of $f(R)$ gravity relevant for our analysis. We begin by recalling the action and field equations, highlighting the role of the extra scalar degree of freedom (the ``scalaron''). We then specialize to a flat FLRW background and derive the modified Friedmann equation in terms of the normalized Hubble parameter $E(z) \equiv H(z)/H_0$. Finally, we outline the main viability conditions for $f(R)$ models and introduce the two benchmark parameterizations used in this work: Hu--Sawicki and Starobinsky.

\subsection{\texorpdfstring{$f(R)$ Gravity Basics}{f(R) Gravity Basics}}

Einstein's General Relativity (GR) is formulated from the Einstein--Hilbert action
\begin{equation}
S_{\rm EH} = \frac{1}{16\pi G} \int d^4x \, \sqrt{-g} \, R + S_m ,
\end{equation}
where $G$ is Newton’s constant, $g = \det(g_{\mu\nu})$ is the determinant of the metric tensor $g_{\mu\nu}$, $R = g^{\mu\nu}R_{\mu\nu}$ is the Ricci scalar curvature, and $S_m$ denotes the matter action. The factor $\sqrt{-g}$ ensures that the action is invariant under general coordinate transformations. Varying this action with respect to the metric yields Einstein’s field equations, $R_{\mu\nu}-\tfrac{1}{2}R g_{\mu\nu}=8\pi G T_{\mu\nu}$, which are second-order in derivatives of the metric.

In $f(R)$ gravity, this action is generalized by promoting the linear dependence on $R$ to a nonlinear function $f(R)$ ~\cite{Sotiriou_2010,Felice_2010,Peralta_2020}:
\begin{equation}
S = \frac{1}{16\pi G} \int d^4x \, \sqrt{-g} \, f(R) + S_m .
\end{equation}
Here $f(R)$ is an arbitrary but differentiable function of the Ricci scalar. This extension is motivated by the idea that higher-order curvature terms naturally appear in the effective action of quantum gravity or string theory, and that such modifications may account for early- and late-time cosmic acceleration without introducing an explicit cosmological constant.

The variation of this generalized action with respect to the metric leads to the field equations
\begin{equation}
f_R R_{\mu\nu} - \tfrac{1}{2} f g_{\mu\nu} + \big(g_{\mu\nu}\Box - \nabla_\mu\nabla_\nu\big) f_R = 8\pi G T_{\mu\nu},
\end{equation}
where $f_R \equiv df/dR$ denotes the first derivative of $f(R)$ with respect to $R$, $\Box \equiv g^{\mu\nu}\nabla_\mu\nabla_\nu$ is the d’Alembertian operator, and $\nabla_\mu$ is the covariant derivative compatible with the metric. Compared to GR, these equations contain higher derivatives of the metric through $\nabla_\mu\nabla_\nu f_R$, which generally  make $f(R)$ theories fourth order in the  metric. The new terms encode the dynamics of an additional scalar degree of freedom hidden inside the metric.

Taking the trace of the above equation gives
\begin{equation}
3\Box f_R + f_R R - 2f = 8\pi G T ,
\end{equation}
where $T = g^{\mu\nu}T_{\mu\nu}$ is the trace of the energy--momentum tensor. This equation is absent in GR and reveals the extra dynamical scalar mode, often called the ``scalaron.’’ This scalar mediates deviations from Einstein gravity and can be viewed as a new field whose mass and dynamics depend on the background curvature through $f_R$ and $f_{RR}$.

In a spatially flat Friedmann--Lemaître--Robertson--Walker (FLRW) spacetime,
\begin{equation}
ds^2 = -dt^2 + a^2(t)\,d\vec{x}^2 ,
\end{equation}
with scale factor $a(t)$ and Hubble parameter $H=\dot{a}/a$, the field equations reduce to modified Friedmann equations. The first Friedmann equation takes the form
\begin{equation}
3 f_R H^2 = 8\pi G \rho_m + \tfrac{1}{2}(f_R R - f) - 3H\dot{f_R} ,
\end{equation}
where $\rho_m$ is the matter density, $f_R \equiv df/dR$, and the derivative of $f_R$ is
\begin{equation}
\dot{f_R} = f_{RR} \dot{R}, \qquad f_{RR} \equiv \frac{d^2 f}{dR^2}.
\end{equation}

The Ricci scalar in this background is
\begin{equation}
R = 6\big(2H^2 + \dot{H}\big) ,
\end{equation}
Both $f(R)$ and its derivatives $f_R$ and $f_{RR}$ couple directly to the expansion rate of the universe and its time evolution.

\section*{Nondimensionalizing and Reducing $f(R)$ Friedmann Equation in Redshift Space}

For cosmological applications, it is convenient to rewrite the background equations of $f(R)$ gravity in terms of the redshift $z$ rather than cosmic time $t$. Recall that the redshift is related to the scale factor by $1+z = a_0/a$, where we take $a_0=1$ today. Since $dz/dt = -(1+z)H$, time derivatives can be replaced by
\begin{equation}
\frac{d}{dt} = -(1+z)H \frac{d}{dz}, 
\qquad 
\dot{H} = -(1+z)H \frac{dH}{dz}.
\end{equation}
This substitution is useful because many observational quantities in cosmology are directly expressed as functions of redshift.

Introducing the normalized Hubble parameter
\begin{equation}
E(z) \equiv \frac{H(z)}{H_0},
\end{equation}
with $H_0$ the present-day Hubble constant, we can express the Ricci scalar as a function of $z$. Starting from $R = 6(2H^2+\dot{H})$ and using the redshift relations above, one obtains
\begin{equation}
R(z) = 6H_0^2 \left[ 2E^2(z) - (1+z)E(z)\frac{dE}{dz} \right].
\end{equation}

At this stage, it is natural to nondimensionalize by defining
\begin{equation}
\mathcal{R}(z) \equiv \frac{R(z)}{H_0^2} 
= 6\left[ 2E^2(z) - (1+z)E(z)E'(z)\right],
\label{eq:ricci_dimless}
\end{equation}
so that the Ricci scalar is expressed entirely in terms of the normalized Hubble function $E(z)$ and its derivative. Differentiating gives
\begin{equation}
\mathcal{R}'(z) = \frac{d\mathcal{R}}{dz} 
= 6\left[ 4E(z)E'(z) - (1+z)(E'(z))^2 - (1+z)E(z)E''(z) \right].
\label{eq:ricciprime_dimless}
\end{equation}

To fully nondimensionalize the field equations, we also rescale the $f(R)$ function:
\begin{equation}
\tilde f(\mathcal{R}) \equiv \frac{f(R)}{H_0^2}, 
\qquad 
\tilde f_{\mathcal{R}} \equiv \frac{df}{d\mathcal{R}}, 
\qquad 
\tilde f_{\mathcal{R}R} \equiv H_0^2 \frac{d^2 f}{d\mathcal{R}^2}.
\end{equation}

With these definitions, the modified Friedmann equation takes the compact, dimensionless form
\begin{equation}
\tilde f_{\mathcal{R}} E^2(z) 
= \Omega_{m0}(1+z)^3 
+ \frac{1}{6}\Big(\tilde f_{\mathcal{R}} \mathcal{R}(z) - \tilde f(\mathcal{R})\Big) 
+ (1+z)\,\tilde f_{\mathcal{RR}}\,E(z)\,\mathcal{R}'(z).
\label{eq:frfriedmann_dimless}
\end{equation}

Here $\Omega_{m0}$ dimensionless matter density parameter. Equation~\eqref{eq:frfriedmann_dimless} is the \textbf{master equation} of $f(R)$ cosmology: a second-order, nonlinear differential equation for the normalized Hubble parameter $E(z)$. Its explicit dependence on $\mathcal{R}(z)$ and $\mathcal{R}'(z)$ encodes the higher-order curvature contributions through $\tilde f_{\mathcal{R}}$ and $\tilde f_{\mathcal{RR}}$.

Writing the equations in terms of redshift $z$ ensures direct compatibility with observations, since distance measures, growth data, and CMB constraints are all expressed in $z$. Nondimensionalization removes explicit factors of $H_0$, leading to numerically stable equations and reducing parameter degeneracies. Together, these steps produce the compact dimensionless form in Eq.~\eqref{eq:frfriedmann_dimless}, which can be solved for any chosen $f(\mathcal{R})$ model and directly confronted with cosmological data.

General Relativity is recovered in the special case $f(\mathcal{R}) = \mathcal{R} - 2\tilde{\Lambda}$, where $\tilde{\Lambda}= \Lambda/H_0^2$ denotes the effective cosmological constant representing the vacuum energy density of space. In this limit, $\tilde{f}_{\mathcal{R}} = 1$ and $\tilde{f}_{\mathcal{RR}} = 0$, so that Eq.~\eqref{eq:frfriedmann_dimless} reduces to the standard $\Lambda$CDM Friedmann equation. In more general cases, however, the evolution of the scale factor is governed not only by the matter content but also by the effective scalar field emerging from curvature corrections. Depending on the functional form of $f(R)$, this scalar can mimic dark energy at late times or drive inflation in the early universe. Solving Eq.~\eqref{eq:frfriedmann_dimless} therefore yields the complete background expansion history, which can be directly confronted with cosmological observations such as supernovae luminosity distances, baryon acoustic oscillations, and anisotropies in the cosmic microwave background.

\subsection{\texorpdfstring{Viable $f(R)$ Models}{Viable f(R) Models}}\label{viable}

A number of functional forms of $f(R)$ have been proposed in the literature, but only a few satisfy the basic requirements of viability~\cite{Sotiriou_2009,Capozziello_2010}. These conditions can be conveniently expressed in the nondimensional variables
\[
\mathcal{R} \equiv \frac{R}{H_0^2}, 
\qquad 
\tilde f(\mathcal{R}) \equiv \frac{f(R)}{H_0^2}, 
\qquad 
\tilde f_{\mathcal{R}} \equiv \frac{df}{dR}, 
\qquad 
\tilde f_{\mathcal{RR}} \equiv H_0^2 \frac{d^2 f}{dR^2},
\]
so that all relations are expressed in terms of $\mathcal{R}$ and its derivatives \cite{Sotiriou_2009,Felice_2010}.The viability conditions are:

\begin{enumerate}
\item[(i)] \textbf{Stable late-time de Sitter attractor:}  
At a constant-curvature fixed point $\mathcal{R} = \mathcal{R}_{\rm dS}$, stability requires
\begin{equation}
\mathcal{R}_{\rm dS} \,\tilde f_{\mathcal{R}}(\mathcal{R}_{\rm dS}) = 2 \tilde f(\mathcal{R}_{\rm dS}), 
\qquad 
0 < \frac{\tilde f_{\mathcal{R}}(\mathcal{R}_{\rm dS})}{\mathcal{R}_{\rm dS}\,\tilde f_{\mathcal{RR}}(\mathcal{R}_{\rm dS})} < 1.
\end{equation}
This ensures the existence of a stable de Sitter solution describing late-time cosmic acceleration.

\item[(ii)] \textbf{Recovery of the matter-dominated era:}  
In the high-curvature regime $\mathcal{R} \gg 1$, the theory must reduce to General Relativity. This requires
\begin{equation}
\lim_{\mathcal{R} \to \infty} \frac{\tilde f(\mathcal{R})}{\mathcal{R}} \to 1, 
\qquad 
\lim_{\mathcal{R} \to \infty} \tilde f_{\mathcal{R}}(\mathcal{R}) \to 1, 
\qquad 
\lim_{\mathcal{R} \to \infty} \mathcal{R}^2 \tilde f_{\mathcal{RR}}(\mathcal{R}) \to 0.
\end{equation}

\item[(iii)] \textbf{Consistency with local gravity tests:}  
In high-density environments ($\mathcal{R} \gg 1$), the scalaron mass
\begin{equation}
m^2(\mathcal{R}) = \frac{H_0^2}{3}\left(\frac{\tilde f_{\mathcal{R}}(\mathcal{R})}{\tilde f_{\mathcal{RR}}(\mathcal{R})} - \mathcal{R}\right)
\end{equation}
must be large so that deviations from GR are exponentially suppressed via the chameleon mechanism.

\item[(iv)] \textbf{Absence of instabilities:}  
The scalar degree of freedom must be ghost-free and tachyon-free, requiring
\begin{equation}
\tilde f_{\mathcal{R}}(\mathcal{R}) > 0, 
\qquad 
\tilde f_{\mathcal{RR}}(\mathcal{R}) > 0,
\end{equation}
which guarantee that the effective gravitational coupling is positive and the scalaron has a real, positive mass.
\end{enumerate}

Taken together, these criteria restrict the form of viable $f(R)$ functions to a narrow class that is consistent with both cosmological and solar-system constraints. Two benchmark examples are the \textbf{Hu–Sawicki model} and the \textbf{Starobinsky model}, which both reproduce the $\Lambda$CDM expansion history at high redshift while introducing small, testable deviations at late times.

\subsection{Benchmark Models}

Among the various functional forms of $f(R)$, two benchmark models have been particularly influential: the Hu--Sawicki model and the Starobinsky model. Both are designed to reproduce the $\Lambda$CDM expansion history at high redshift while introducing small deviations at late times that can be constrained by cosmological observations. For consistency with the dimensionless master equation, we express these models in terms of nondimensional variables as defined in Section \ref{viable}.

\subsubsection{The Hu--Sawicki Model}

\subsubsection*{The Hu--Sawicki Model}

The Hu--Sawicki (HS) model~\cite{Hu_2007} represents one of the most widely studied and observationally viable formulations of $f(R)$ gravity. It is constructed with the explicit goal of recovering General Relativity (GR) in the high-curvature regime, corresponding to the early Universe and dense astrophysical environments, while allowing \textit{small but measurable deviations at late times} when the curvature drops to cosmological scales. 

In nondimensional form, the model can be written as
\begin{equation}
\tilde f(\mathcal{R}) = \mathcal{R} - 2\tilde\Lambda \left( 1 - \frac{1}{1 + (\mathcal{R}/b\tilde\Lambda)^n} \right),
\label{eq:HS_dimless}
\end{equation}
where $\tilde\Lambda = \Lambda/H_0^2$ is the effective cosmological constant in units of $H_0^2$, $b$ is a dimensionless deviation parameter, and $n$ is a positive integer controlling the steepness of the transition between the GR and modified-gravity regimes. 

Physically, the parameter $b$ quantifies \textit{how strongly the model departs from $\Lambda$CDM}: as $b \to 0$, the correction term in Eq.~\eqref{eq:HS_dimless} vanishes, and one recovers the standard GR limit $\tilde f(\mathcal{R}) = \mathcal{R} - 2\tilde\Lambda$. For finite $b$, however, the additional curvature-dependent term modifies the effective gravitational coupling at low curvature, allowing the late-time cosmic acceleration to emerge from \textit{geometry itself rather than from an explicit cosmological constant}. This delicate balance---between preserving GR in high-density regions and reproducing accelerated expansion at large scales---makes the Hu--Sawicki model a benchmark framework for testing modified gravity in the precision-cosmology era.

For the commonly used $n=1$ case, the derivatives are
\begin{align}
\tilde f_{\mathcal{R}} &= 1 - \frac{2\tilde\Lambda}{b\tilde\Lambda + \mathcal{R}}, \\
\tilde f_{\mathcal{RR}} &= \frac{2\tilde\Lambda}{(b\tilde\Lambda + \mathcal{R})^2}.
\end{align}
These enter directly into the dimensionless master equation \eqref{eq:frfriedmann_dimless}, providing a two-parameter deformation of $\Lambda$CDM characterized by $(\tilde\Lambda, b)$.

\subsubsection{The Starobinsky Model}

The Starobinsky model~\cite{Starobinsky_1980} is another cornerstone of viable $f(R)$ cosmology, originally proposed as an inflationary model and later adapted to explain late-time acceleration through geometric corrections to Einstein's theory. In contrast to the Hu--Sawicki formulation, which introduces a dimensionless deformation parameter, the Starobinsky model modifies gravity through a \textit{dimensional curvature scale} that governs when deviations from General Relativity (GR) become significant. Specifically, the model introduces corrections that are negligible at high curvature---thereby preserving GR in the early Universe and within dense astrophysical systems---but become important at low curvature, where cosmic acceleration sets in. 

In nondimensional variables and for $n=1$, the function can be expressed as
\begin{equation}
\tilde f(\mathcal{R}) = \mathcal{R} - 2\tilde\Lambda \left( 1 - \frac{1}{1 + (\mathcal{R}^2/\mathcal{R}_c^2)} \right),
\label{eq:Star_dimless}
\end{equation}
where $\tilde\Lambda = \Lambda/H_0^2$ is the effective cosmological constant, and $\mathcal{R}_c = R_c/H_0^2$ defines the characteristic curvature scale that separates the GR-like high-curvature regime from the modified-gravity low-curvature regime. 

Physically, $\mathcal{R}_c$ determines \textit{where} in curvature space the modifications to GR become relevant: when $\mathcal{R} \gg \mathcal{R}_c$, the correction term in Eq.~\eqref{eq:Star_dimless} becomes negligible, and the model effectively reproduces GR; when $\mathcal{R} \sim \mathcal{R}_c$, curvature-driven effects dominate and can account for late-time cosmic acceleration without invoking a large cosmological constant. In the limit $\mathcal{R}_c \to 0$, the modification term vanishes, recovering $\tilde f(\mathcal{R}) = \mathcal{R} - 2\tilde\Lambda$, i.e.\ the $\Lambda$CDM scenario. The presence of this curvature threshold makes the Starobinsky model particularly elegant, as it geometrically links the onset of cosmic acceleration to the evolution of spacetime curvature itself.

The derivatives are
\begin{align}
\tilde f_{\mathcal{R}} &= 1 - \frac{4\tilde\Lambda\,\mathcal{R}/\mathcal{R}_c^2}{\left(1+\mathcal{R}^2/\mathcal{R}_c^2\right)^2}, \\
\tilde f_{\mathcal{RR}} &= \frac{4\tilde\Lambda(\mathcal{R}_c^2 - 3\mathcal{R}^2)}{\mathcal{R}_c^4\left(1+\mathcal{R}^2/\mathcal{R}_c^2\right)^3}.
\end{align}
 These enter directly into the dimensionless master equation \eqref{eq:frfriedmann_dimless}, providing a two-parameter deformation of $\Lambda$CDM characterized by $(\tilde\Lambda, \mathcal{R}_c )$.

\begin{figure}[ht]
  \centering
  \resizebox{\textwidth}{!}{%
    \begin{tikzpicture}[]
      \node (master) [eqbox, align=left] {
        \textbf{Master Equation (dimensionless):} \\[0.5em]
        {\Large $
          F(E, E', E''; z, \Omega_{m0}, \theta ) =
          \tilde f_{\mathcal{R}} E^2(z)
          - \Omega_{m0}(1+z)^3
          + \tfrac{1}{6}\big(\tilde f_{\mathcal{R}} \mathcal{R}(z) - \tilde f(\mathcal{R})\big)
          + (1+z)\tilde f_{\mathcal{RR}}E(z)\mathcal{R}'(z)=0
        $}
      };

      \node (defs) [eqbox, align=left, below=of master] {
        \textbf{Dimensionless curvature:} \\[0.5em]
        $\mathcal{R}(z) = \dfrac{R(z)}{H_0^2} = 6\!\left[2E^2(z) - (1+z)E(z)E'(z)\right]$ \\[0.5em]
        $\mathcal{R}'(z) = 6\!\left[4E(z)E'(z) - (1+z)(E'(z))^2 - (1+z)E(z)E''(z)\right]$
      };

      \draw[arrow] (master) -- (defs);

      \node (HS) [modelbox, align=left, below left=2.8cm and -1.2cm of defs] {
        \textbf{Hu--Sawicki (For $n=1$) \& $\theta=(b, \tilde\Lambda)$} \\[0.5em]
        $\tilde f(\mathcal{R}) = \mathcal{R} - 2\tilde\Lambda \!\left(1-\dfrac{1}{1+\mathcal{R}/(b\tilde\Lambda)}\right)$ \\[0.5em]
        $\tilde f_{\mathcal{R}} = 1 - \dfrac{2\tilde\Lambda}{b\tilde\Lambda+\mathcal{R}}$ \\[0.5em]
        $\tilde f_{\mathcal{RR}} = \dfrac{2\tilde\Lambda}{(b\tilde\Lambda+\mathcal{R})^2}$, where
        $\tilde\Lambda= \dfrac{\Lambda}{H_0^2}$
      };

      \node (Star) [modelbox, align=left, below right=2.8cm and -1.2cm of defs] {
        \textbf{Starobinsky (For $n=1$ \& $\theta=(\mathcal{R}_c, \tilde\Lambda)$)} \\[0.5em]
        $\tilde f(\mathcal{R}) = \mathcal{R} - 2\tilde\Lambda \!\left(1-\dfrac{1}{1+\mathcal{R}^2/\mathcal{R}_c^2}\right)$ \\[0.5em]
        $\tilde f_{\mathcal{R}} = 1 - \dfrac{4\tilde\Lambda \,\mathcal{R}/\mathcal{R}_c^2}{(1+\mathcal{R}^2/\mathcal{R}_c^2)^2}$ \\[0.5em]
        $\tilde f_{\mathcal{RR}} = \dfrac{4\tilde\Lambda(\mathcal{R}_c^2-3\mathcal{R}^2)}{\mathcal{R}_c^4(1+\mathcal{R}^2/\mathcal{R}_c^2)^3}$, where $\tilde{\mathcal{R}}_c= \dfrac{R_c}{H_0^2}$
      };

      \node (LCDM) [modelbox, align=left, below=4cm of defs] {
        \textbf{$\Lambda$CDM (limit)} \\[0.5em]
        $\tilde f(\mathcal{R}) = \mathcal{R} - 2\tilde\Lambda$,\\
        $\tilde f_{\mathcal{R}} = 1$,\\
        $\tilde f_{\mathcal{RR}} = 0$
      };

      \draw[arrow] (defs.south west) -- ++(-1.0,-0.5) |- (HS.north);
      \draw[arrow] (defs.south east) -- ++(1.0,-0.5) |- (Star.north);
      \draw[arrow] (defs.south) -- ++(0,-0.5) |- (LCDM.north);

    \end{tikzpicture}%
  } 

\caption{\textbf{Equation Flow for $f(R)$ Cosmology (Dimensionless Form):} Flowchart summarizing the dimensionless master equation of $f(R)$ cosmology in redshift space and its related quantities ($\mathcal{R}$, $\mathcal{R}'$, $\tilde f$, $\tilde f_{\mathcal{R}}$, $\tilde f_{\mathcal{RR}}$) for the Hu--Sawicki and Starobinsky models. For a given matter density $\Omega_{m0}$ and model parameters $\theta$, this nonlinear equation can be solved by determining $E(z)$ together with its derivatives $E'(z)$ and $E''(z)$. Hence, solving Eq.~\eqref{eq:frfriedmann_dimless} is the primary aim of background cosmological analysis in $f(R)$ gravity.}
 \label{fig:flowchart1}
\end{figure}
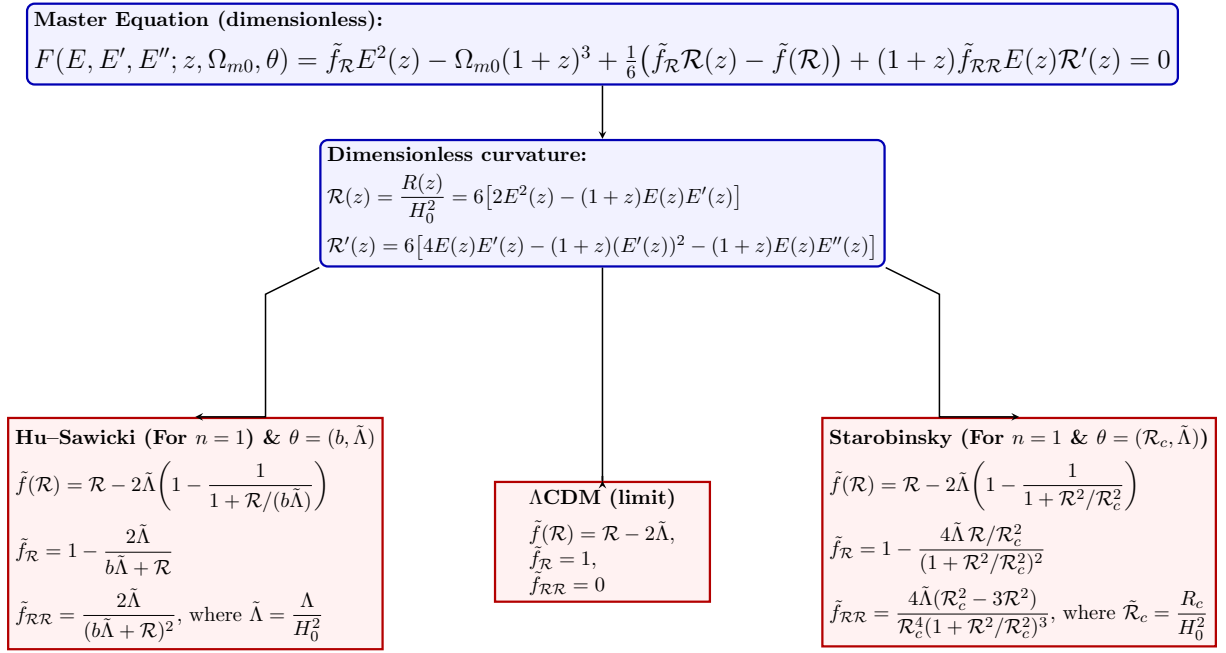

It is tempting to note that the Hu--Sawicki and Starobinsky functions exhibit a similar rational form and may appear equivalent under the parameter identification $b\tilde{\Lambda} \simeq \mathcal{R}_c$ and $n=2$. However, this correspondence is only superficial. The two models differ in the power of curvature entering their correction terms: the Hu-Sawicki model introduces a $(\mathcal{R}/b \tilde{\Lambda})^{-n}$ suppression, whereas the Starobinsky model employs a $(\mathcal{R}^2/\mathcal{R}_c^2)^{-n}$ dependence. As a consequence, their first and second derivatives, $f_{\mathcal{R}}$ and $f_{\mathcal{RR}}$, evolve differently with curvature, leading to distinct scalaron masses, stability conditions, and screening behaviors. Even when tuned to similar numerical scales, the two models occupy different regions of the $(\mathcal{R}, f_{\mathcal{R}}, f_{\mathcal{RR}})$ phase space and generate inequivalent expansion histories. Hence, they are best regarded as two structurally analogous but physically distinct realizations of viable $f(R)$ gravity. The flow of the master equation and the respective solutions for the benchmark models are summarized schematically in Fig.~\ref{fig:flowchart1}.

\section{Numerical Methods}
\label{sec:Numerical_3}

The modified Friedmann equation of $f(R)$ gravity, expressed in redshift space (see Figure.~\eqref{fig:flowchart1}), takes the form of a nonlinear, second--order ODE in the normalized Hubble function $E(z) = H(z)/H_0$. Unlike $\Lambda$CDM, where the expansion history follows from a closed algebraic formula, the $f(R)$ case couples $E(z)$ to its first and second derivatives through the Ricci scalar $R(z)$ and its redshift derivative $dR/dz$.  This coupling makes the system \emph{stiff}. In practice, stiffness means that the differential equation contains both slowly varying modes (set by the overall Hubble expansion) and rapidly varying modes (introduced by the higher derivatives of $E(z)$ through $\mathcal{R}$ and $d\mathcal{R}/dz$). Near the $\Lambda$CDM limit, where the deviation parameters are small, these rapid modes are especially pronounced: even tiny changes in the input parameters can produce sharp variations in $E'(z)$ and $E''(z)$. Standard ODE solvers such as explicit Runge--Kutta or Bulirsch--Stoer then struggle, because numerical stability forces them to adopt extremely small step sizes. This makes the integration expensive and often unstable, with solutions either diverging or failing to converge altogether \cite{Hu_2007,Tsujikawa_2008,Amendola_2007}.The stiffness problem is therefore one of the principal obstacles in obtaining reliable background solutions for $f(R)$ cosmologies.

A further difficulty with solving the background equations of $f(R)$ gravity is that the system is second order in $E(z)$ but only one clear boundary condition is available: the normalization $E(0)=1$, which ensures $H(z=0)=H_0$.  In contrast to ordinary differential equations where two boundary conditions fix a unique solution, here the second condition is not obvious.  One often imposes an asymptotic condition in the high-redshift regime, e.g.\ requiring the solution to approach the $\Lambda$CDM form as $z \to \infty$, but numerically enforcing such a condition is unstable.  This underdetermination makes direct integration of the system even more challenging and motivates approximate schemes such as perturbative expansions \cite{Basilakos_2013,Odintsov_2026,Nunes_2017}.

To mitigate the stiffness problem, a commonly adopted strategy is the perturbative b-expansion proposed by Basilakos et al.~\cite{Basilakos_2013,Bamba_2012}. In this approach, the $f(R)$ background evolution is expressed as a small deviation from the standard $\Lambda$CDM expansion history, allowing the equations to be solved analytically to leading order. This formulation is computationally efficient because it avoids direct integration of the stiff differential system. Nevertheless, its accuracy deteriorates as the deviation parameters increase, which coincides with the parameter space most relevant for probing observable signatures of modified gravity.

To overcome these difficulties, we adopt a \emph{global spectral method} based on Chebyshev polynomials, which avoids stepwise propagation and instead constructs the solution across the full redshift interval in one shot.  Let us develop the mathematical understanding of the proposed method:

\subsection{\texorpdfstring{Chebyshev Expansion of $E(z)$ and Methodology}{Chebyshev Expansion of E(z) and Methodology}}\label{cheby}

Under the spectral approximation,  we represent the expansion history by a truncated Chebyshev series :

\begin{equation}
E(z) \approx \sum_{k=0}^N a_k \, T_k\!\big(x(z)\big),
\label{tcs}
\end{equation}

\noindent where $T_k(x)$ are Chebyshev polynomials of the first kind, $a_k$ are unknown expansion coefficients, and $x(z)$ is a linear transformation that maps the physical redshift interval $z \in [0,z_{\max}]$ to the canonical Chebyshev domain $x \in [-1,1]$:

\begin{equation}
x(z) = \frac{2z}{z_{\max}} - 1.
\end{equation}

This mapping ensures that the entire expansion history, from today ($z=0$) to the high-redshift matter-dominated regime, is covered by the Chebyshev approximation. This guarantees that the Chebyshev expansion in $x$ corresponds to an optimally convergent polynomial representation of $E(z)$ in the physical redshift interval.\\

For our purposes, Chebyshev polynomials are useful because of their properties. 
Chebyshev polynomials are almost optimal for global polynomial approximation, as they minimize the maximum interpolation error, a property known as the \emph{Chebyshev minimax property}. The non-uniform distribution of nodes suppresses the Runge phenomenon, which otherwise leads to oscillations in naive polynomial fits with equally spaced nodes. Moreover, for smooth functions such as $E(z)$, a Chebyshev expansion converges exponentially, which means that high precision can be achieved with relatively few terms $N$ ~\cite{Boyd:2001Spectral,Trefethen:2000Spectral,Shen:2011Spectral}. \\

Here, we use the collocation method to solve the ordinary differential equations (ODEs) \ref{eq:master_cheb}. In this numerical method, the solution is approximated so that the equation is exactly satisfied only at a limited number of points in the domain, called \emph{collocation nodes}. Instead of enforcing the equation everywhere, we apply it only at these selected points. This converts the continuous problem into a smaller system of algebraic equations that can be solved on a computer. The accuracy of the method mainly depends on where these nodes are placed and on the order of the functions used to approximate the solution.\\

In this work, we apply the \emph{collocation method} using \emph{Chebyshev polynomials} as the basis functions. In Chebyshev spectral methods, the collocation nodes are chosen as the \emph{Chebyshev--Gauss--Lobatto points}~\cite{Grandcl_2009,Hejranfar_2015,Galal_2016,Karjanto_2020}, which are defined as
\begin{equation}
x_j = \cos\!\left(\frac{\pi j}{N}\right), \qquad j = 0, 1, \dots, N.
\end{equation}

which correspond to the extremal points of $T_N(x)$.  Mapped back to redshift space, the nodes are

\begin{equation}
z_j = \frac{z_{\max}}{2}\,(x_j+1).
\label{zj}
\end{equation}

It is important to note that these nodes are not the zeros of $E(z)$, but special points where we make the modified Friedmann equation hold exactly. The equation is applied only at these discrete points, while between them the solution is smoothly determined by the overall Chebyshev polynomial. This approach keeps the errors extremely small and ensures that the solution changes smoothly across the domain. Because the nodes include both ends of the interval, they give very good numerical stability and fast convergence for smooth functions, allowing an accurate and compact representation of the continuous problem.\\

At the Chebyshev--Gauss--Lobatto collocation points ($z_j$), and for any fixed values of $\Omega_{m0}$ and model parameters $\theta$ (e.g.\ $b$, $\mathcal{R}_c$), the dimensionless master equation is enforced provided $E(z)$, $E'(z)$, and $E''(z)$ are reconstructed from the truncated Chebyshev series of Eq.~\eqref{tcs}:

\begin{equation}
    F(E,E',E''; z, \Omega_{m0}, \theta ) =
    \tilde f_{\mathcal{R}} E^2(z) 
    - \Omega_{m0}(1+z)^3  
    + \tfrac{1}{6}\Big(\tilde f_{\mathcal{R}} \mathcal{R}(z) - \tilde f(\mathcal{R})\Big)  
    + (1+z)\tilde f_{\mathcal{RR}}E(z)\mathcal{R}'(z) = 0. 
    \label{eq:master_cheb}
\end{equation}

Here the dimensionless Ricci scalar $\mathcal{R}(z)$ and its derivative $\mathcal{R}'(z)$ depend explicitly on $E(z), E'(z)$, and $E''(z)$. These derivatives are obtained spectrally from the Chebyshev expansion using differentiation matrices. At the collocation nodes $x_i$ in the canonical domain $[-1,1]$, one has

\begin{equation}
E'(x_i) = \sum_{j=0}^N D^{(1)}_{ij} \, E(x_j), 
\qquad
E''(x_i) = \sum_{j=0}^N D^{(2)}_{ij} \, E(x_j),
\label{eq:cheb_diff}
\end{equation}

where $D^{(1)}$ and $D^{(2)}$ are the standard Chebyshev differentiation matrices of order one and two, respectively.  
Since the ODE is written in redshift $z \in [0,z_{\max}]$ rather than in the canonical Chebyshev domain, the mapping

\begin{equation}
x(z) = \frac{2z}{z_{\max}} - 1
\end{equation}

implies a Jacobian rescaling of the differentiation matrices:

\begin{equation}
D^{(1)}_z = \frac{2}{z_{\max}} \, D^{(1)}, 
\qquad
D^{(2)}_z = \left(\frac{2}{z_{\max}}\right)^2 D^{(2)}.
\label{eq:rescaled_diff}
\end{equation}

This rescaling guarantees that $E'(z)$ and $E''(z)$ are evaluated at machine precision, without resorting to finite-difference approximations. Thus, the Chebyshev collocation approach transforms the nonlinear second-order differential equation Eq.~\eqref{eq:master_cheb} into a set of coupled algebraic equations at the nodes $z_j$, which can be solved for the spectral coefficients of $E(z)$.

\subsection{Solving the System of Coupled Nonlinear Algebraic Equations}

The truncated Chebyshev expansion of the normalized Hubble parameter $E(z)$,given in Eq.~\eqref{tcs}, introduces $N+1$ unknown coefficients $\{a_k\}_{k=0}^N$. By substituting this expansion into the dimensionless master equation Eq.~\eqref{eq:master_cheb}, and enforcing it at the $N+1$ Chebyshev--Gauss--Lobatto collocation nodes $z_j$ defined by  Eq.~\eqref{zj}, we obtain a closed system of $N+1$ nonlinear algebraic equations:

\begin{equation}
F_i(\{a_k\}; z_j, \Omega_{m0}, \theta) = 0, 
\qquad i = 0,1,\dots,N,
\end{equation}

where $\theta$ denotes the model parameters (e.g.\ $b$, $\mathcal{R}_c$).  In this way, the original stiff second-order ODE for $E(z)$ has been replaced by a coupled system of nonlinear algebraic equations for the spectral coefficients $\{a_k\}$.  \textit{This is the central idea of our collocation method: transforming a differential problem into an algebraic one.}

\paragraph{Numerical solution strategy.}
The collocation method converts the original differential equation into a nonlinear system $\vec{F}({a_k})=0$ for the Chebyshev coefficients. Such systems are high-dimensional and stiff, making simple root-finding unreliable. 

Newton--Raphson converges quadratically near the true root \cite{Press:2007NumRecipes}, but requires Jacobians, is highly sensitive to initial guesses, and often diverges for ill-conditioned problems. Newton--Krylov methods \cite{knoll_2004} avoid explicit Jacobians and work well for very large sparse systems, but for the moderate sizes here ($N \lesssim 50$) they introduce unnecessary overhead. Both proved useful for benchmarking but were either unstable (Newton--Raphson) or inefficient (Newton--Krylov) for $f(R)$ cosmology. 

\paragraph{Powell--hybr method.}  
Given these limitations, we adopted the Powell--hybr algorithm \cite{More:1980Hybrid}, which proved to be both robust and efficient across the parameter ranges we studied. This method blends two strategies: steepest descent, which guarantees global stability, and Newton’s method, which gives rapid convergence near the solution. The core idea is a \emph{trust-region framework} that adaptively limits the size of each step depending on how well the quadratic model predicts the reduction in the residual. This allows the solver to remain stable even when starting from a poor initial guess, while still recovering Newton-like speed close to the solution.  

For reproducibility, we used the implementation available in \texttt{scipy.optimize.root} with \texttt{method="hybr"} \cite{Virtanen_2020}. This routine includes important safeguards such as adaptive trust-region adjustments, finite-difference Jacobian approximations, and automatic monitoring of the residual norm. Using this well-tested library avoids the need to hand-code Jacobian evaluations and ensures that our solver is consistent with widely used numerical standards.

\paragraph{Implementation and algorithm.}  

In practice, the algorithm proceeds as follows:  
\begin{enumerate}
\item Initialize with an approximate solution $\{a_k^{(0)}\}$, obtained by fitting the Chebyshev expansion to the $\Lambda$CDM solution.  
\item Construct an approximate Jacobian (via finite differences or secant updates).  
\item Propose a Newton step by solving $J \Delta a = -\vec{F}$.  
\item Restrict the step $\Delta a$ to a trust region if the Newton update is unreliable.  
\item Adapt the trust-region radius based on the achieved versus predicted reduction in 
$\|\vec{F}\|$.  
\item Repeat steps 2--5 until convergence is reached.  
\end{enumerate}

\begin{landscape} 
\section*{Flowchart of Methodology} 
\begin{figure}
    \centering
    \includegraphics[scale=0.6]{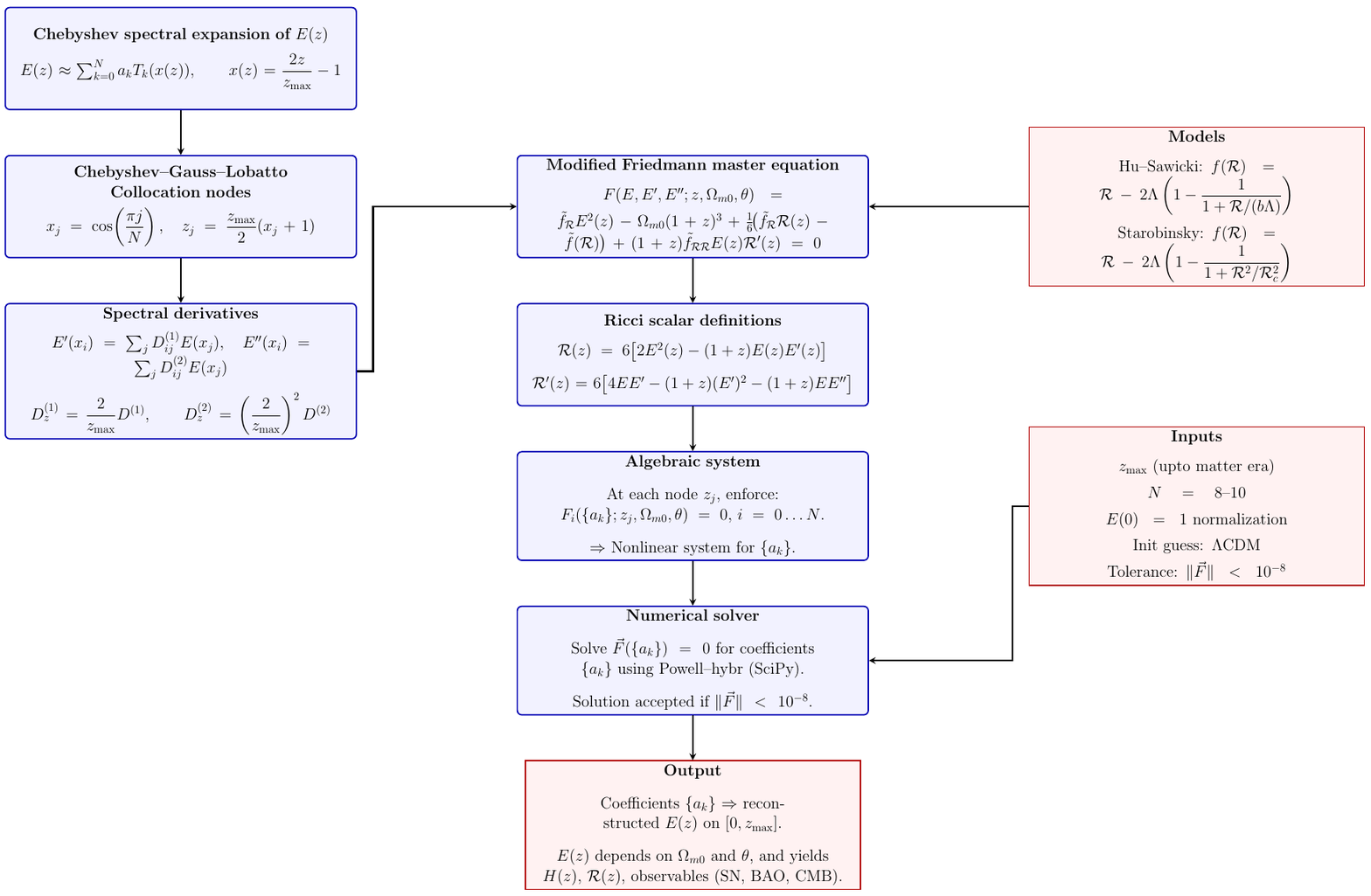}
    \caption{Flowchart of the Chebyshev collocation method for $f(R)$ cosmology, showing the master equation $F(E,E',E'';z,\Omega_{m0},\theta)=0$, Ricci scalar definitions, algebraic system setup, solver, and inputs/models required for coding. Refer to Appendix \ref{app:cheb_solver} for the stepwise process.}
    \label{fig:flowchart}
\end{figure}
\end{landscape}

This adaptive strategy combines the global robustness of gradient methods with the rapid local convergence of Newton’s method, making Powell--hybr particularly well-suited to the stiff, nonlinear algebraic systems that arise in $f(R)$ cosmology.

\subsubsection{User inputs required for the calculation}

Before solving the $f(R)$ problem numerically, several key inputs must be specified by the user. These define the computational domain, the resolution of the Chebyshev expansion, and the numerical setup of the solver:

\begin{enumerate}[label=(\alph*)]
\item \textbf{Choice of $z_{\max}$:}  
The maximum redshift defines the domain $z \in [0,z_{\max}]$ and fixes the mapping to Chebyshev nodes via
\begin{equation}
x(z) = \frac{2z}{z_{\max}} - 1.
\end{equation}
A suitable $z_{\max}$ must be large enough to provide stable boundary conditions, yet not so large as to introduce unnecessary physics (e.g.\ radiation domination).  From our convergence tests, values $z_{\max} \lesssim 10$ lead to unstable or failed solutions, while $z_{\max}=20$--$30$ ensures reliable convergence with residual norms $\|\vec{F}\|\lesssim10^{-12}$ for $z\leq 2$.  We therefore adopt $z_{\max}=30$, which extends the reconstruction safely into the matter-dominated era ($E(z)\propto (1+z)^{3/2}$), stabilizes the Chebyshev expansion against edge effects, and reproduces late-time dynamics with percent-level accuracy.

\item \textbf{Polynomial order $N$:}  
The Chebyshev truncation order $N$ determines the number of spectral coefficients $\{a_k\}$, and hence the resolution of $E(z)$. While larger $N$ in principle improves accuracy, it also raises the condition number of the differentiation matrices, amplifying numerical instabilities.  Our systematic scan shows that very low orders ($N\leq 4$) underresolve the expansion history, while high orders ($N\gtrsim 12$) degrade stability due to ill-conditioning. The optimal regime is $N=8$--$10$, which yields converged solutions with residual norms $\|\vec{F}\|\sim 10^{-12}$ and minimal sensitivity to overfitting.  Thus, even relatively modest truncation orders are sufficient for robust and precise reconstruction of $E(z)$ in $f(R)$ background cosmology.

\item \textbf{Normalization condition:}  
To anchor the solution to physical observables, we impose the condition
\begin{equation}
E(0) = 1,
\end{equation}
which ensures that the present-day Hubble rate is $H(z=0)=H_0$.  This can be enforced either by replacing one collocation equation with $E(0)=1$, or by solving the full system and rescaling the coefficients afterward.  In our implementation, we adopt the latter approach: once the coefficients $\{a_k\}$ are obtained, we normalize the Chebyshev expansion by dividing through its value at $z=0$.  This guarantees the correct normalization while preserving the integrity of the nonlinear system and avoiding additional constraints in the solver.

\item \textbf{Initial guess and convergence tolerance:}  
Efficient convergence of the Powell--hybr solver requires a good initial guess for $\{a_k\}$.  We obtain this by projecting the truncated Chebyshev expansion onto the $\Lambda$CDM background,
which corresponds to the $f(R) = R - 2\Lambda$ limit.  This initialization provides rapid convergence for models close to $\Lambda$CDM while still working robustly for moderate deviations.  The solver iterates until the residual norm satisfies
\begin{equation}
\|\vec{F}(\{a_k\})\| < 10^{-8},
\end{equation}
a threshold chosen to ensure machine-precision reconstruction of the background expansion.  Our convergence scans confirm that this tolerance yields stable and reproducible results across different parameter values.

\end{enumerate}

\section{Observational Data}
\label{sec:data_4}

Constraining $f(R)$ models requires a comparison of the reconstructed background expansion$E(z)$ with cosmological observations. We focus on late--time probes that directly test the Hubble expansion history, and combine them into a joint likelihood framework.

\subsection{\texorpdfstring{Cosmic Chronometer $H(z)$ Data}{Cosmic Chronometer H(z) Data}}


We adopt the most recent compilation of observational Hubble parameter data (OHD), consisting of 32 measurements in the redshift range $0.07 \leq z \leq 1.965$. These measurements are obtained using the \emph{cosmic chronometer} (CC) technique \cite{Jimenez_2002,Moresco_2012,Moresco_2016}, which is among the very few methods capable of determining $H(z)$ in a completely model--independent way.  

Cosmic chronometers are passively evolving early--type galaxies that formed most of their stars at high redshift and have since evolved without further star formation. The differential age method exploits the relative age difference $\Delta t$ between two such populations observed at slightly different redshifts $\Delta z$, yielding
\begin{equation}
H(z) = -\frac{1}{1+z}\frac{\Delta z}{\Delta t}.
\end{equation}
This relation directly ties the cosmic expansion rate to galaxy chronometry, without assuming any underlying cosmological model.  

The robustness of the CC method depends on stellar population synthesis (SPS) modeling, which introduces systematics from metallicity, star--formation history, and SPS library choices. These remain active areas of refinement \cite{Moresco_2016,Jimenez_2023}. Despite these challenges, CC measurements provide unique, direct constraints on $H(z)$, complementary to integral probes such as Type Ia supernovae.

\subsection{Union 3.0 / Unity 3.0 Supernova Sample}  

To complement the $H(z)$ data, we also make use of the latest Union~3.0 compilation of Type~Ia supernovae, analyzed within the UNITY Bayesian framework \cite{Rubin_2025}. This dataset brings together more than 2000 SNe~Ia from 24 different surveys, covering redshifts from the very local Universe ($z\sim0.01$) out to $z>2.26$. All supernova light curves have been consistently re-fitted with the SALT3 model, ensuring that color, shape, and other calibration effects are treated in a uniform way.  

The UNITY framework is designed to account for the many sources of systematic uncertainty that affect supernova cosmology. These include photometric calibration (zeropoints and bandpasses), host-galaxy correlations, peculiar velocities at low redshift, and survey selection effects. By handling all of these within a hierarchical Bayesian approach, the analysis is able to propagate uncertainties more reliably and avoid biases from survey inhomogeneity.  

For cosmological parameter estimation it is often more practical to use a compressed version of this dataset rather than the full light-curve analysis. Rubin et al.\ (2025) therefore provide a set of 22 distance modulus nodes that summarize the information content of the full Union~3.0 sample\footnote{Click here for Union 3.0 dataset \url{https://github.com/CobayaSampler/sn_data} and \url{https://github.com/rubind/union3_release}}. These nodes span the redshift range up to $z\approx 2.26$ (in our compressed dataset) and are accompanied by a full $22\times22$ covariance matrix that captures both statistical errors and systematic correlations. In this way, the Union~3.0 / Unity  dataset  provides the most comprehensive and carefully calibrated SN~Ia Hubble diagram currently available and offers a powerful complement to direct $H(z)$ measurements from cosmic chronometers.  

\section{Result}  
\label{sec:result_5}

To constrain the parameters of $f(R)$ cosmologies, we construct a joint likelihood from cosmic chronometer $H(z)$ measurements and the Union~3.0 supernova dataset. Assuming statistical independence, the likelihood factorizes as $\mathcal{L}_{\rm joint} = \mathcal{L}_{H(z)} \times \mathcal{L}_{\rm SN}$, or equivalently $\chi^2_{\rm joint} = \chi^2_{H(z)} + \chi^2_{\rm SN}$. The $H(z)$ contribution is computed as
\[
\chi^2_{H(z)} = \sum_i \frac{\big(H_0E(z_i) - H^{\rm obs}_i\big)^2}{\sigma_{H,i}^2},
\]
while the supernova term uses the compressed Union~3.0 nodes with full covariance,
\[
\chi^2_{\rm SN} = \vec{r}^{\,T} C^{-1} \vec{r}, 
\qquad 
\vec{r} = \mu_{\rm th}(z_i) - \mu_i ,
\]
where $\mu_{\rm th}(z)$ is obtained from the luminosity distance $d_L(z) = (1+z)\int_0^z dz'/[H_0E(z')]$. In $f(R)$ gravity the function $E(z)$ is not given analytically; instead we reconstruct it spectrally using the truncated Chebyshev series, with coefficients determined by solving the collocation system introduced earlier.  

The likelihood is supplemented by physically motivated priors. For the matter density, we impose a Gaussian prior centered on the Planck-preferred value, $\Omega_{m0} \sim \mathcal{N}(0.3,\,0.02^2)$. Similarly, the cosmological constant is assigned a Gaussian prior, $\tilde{\Lambda} \sim \mathcal{N}(2.0,\,0.5^2)$, where $\tilde{\Lambda} \equiv \Lambda/H_0^2$. For the Hu--Sawicki model, the deviation parameter $b$ follows a uniform prior $10^{-8}<b<1$, with $b<10^{-8}$ treated as numerically equivalent to $\Lambda$CDM. For the Starobinsky model we adopt a log-flat prior on the curvature scale parameter $\mathcal{R}_c \equiv R_c/H_0^2$, uniform in $\log \mathcal{R}_c$ over the range $10^{-6} < \mathcal{R}_c < 10^2$, which ensures equal weight across orders of magnitude. In addition, we restrict parameters to the physically reasonable domains $0.2 < \Omega_{m0} < 0.5$ and $0.3 < \tilde{\Lambda} < 4.5$. Further, a uniform prior was imposed on the Hubble constant, $H_0$, within the range $40 < H_0 < 90$, ensuring an unbiased exploration of the parameter space across all plausible cosmological scales.

This combination of Gaussian priors on well-measured quantities and log-flat priors on scale parameters allows us to remain conservative where uncertainties are large while incorporating external knowledge where the cosmology is already tightly constrained. Together with the $H(z)$ and SN data, this framework provides a consistent approach to parameter estimation in $f(R)$ cosmology.

We explore the parameter space using the affine-invariant ensemble sampler \texttt{emcee} \citep{Foreman_2013}. This MCMC method evolves an ensemble of walkers and is efficient for correlated and anisotropic posteriors, which commonly appear in $f(R)$ cosmology. At each likelihood evaluation, the background expansion $E(z)$ is reconstructed with the Chebyshev collocation method (Sec. \ref{sec:Numerical_3}), so that the model predictions remain fully consistent with the non-perturbative equations.

Since Bayesian inference typically requires tens of thousands of likelihood evaluations, computational speed is crucial. To achieve this, we adopt two practical optimizations. First, we use a \emph{warm-start} strategy: once the solver finds a solution for one parameter set, the resulting Chebyshev coefficients $\{a_k\}$ are used as the initial guess for nearby points. This greatly reduces the number of iterations needed for convergence. Second, we implement caching, storing previously computed solutions and reusing them whenever the sampler proposes the same parameter combination. These steps make the Chebyshev solver both fast and stable, allowing us to run long MCMC chains without relying on approximate perturbative expansions.

\begin{figure}[ht]
    \centering
    \begin{subfigure}[t]{0.48\textwidth}
        \centering
        \includegraphics[width=\textwidth]{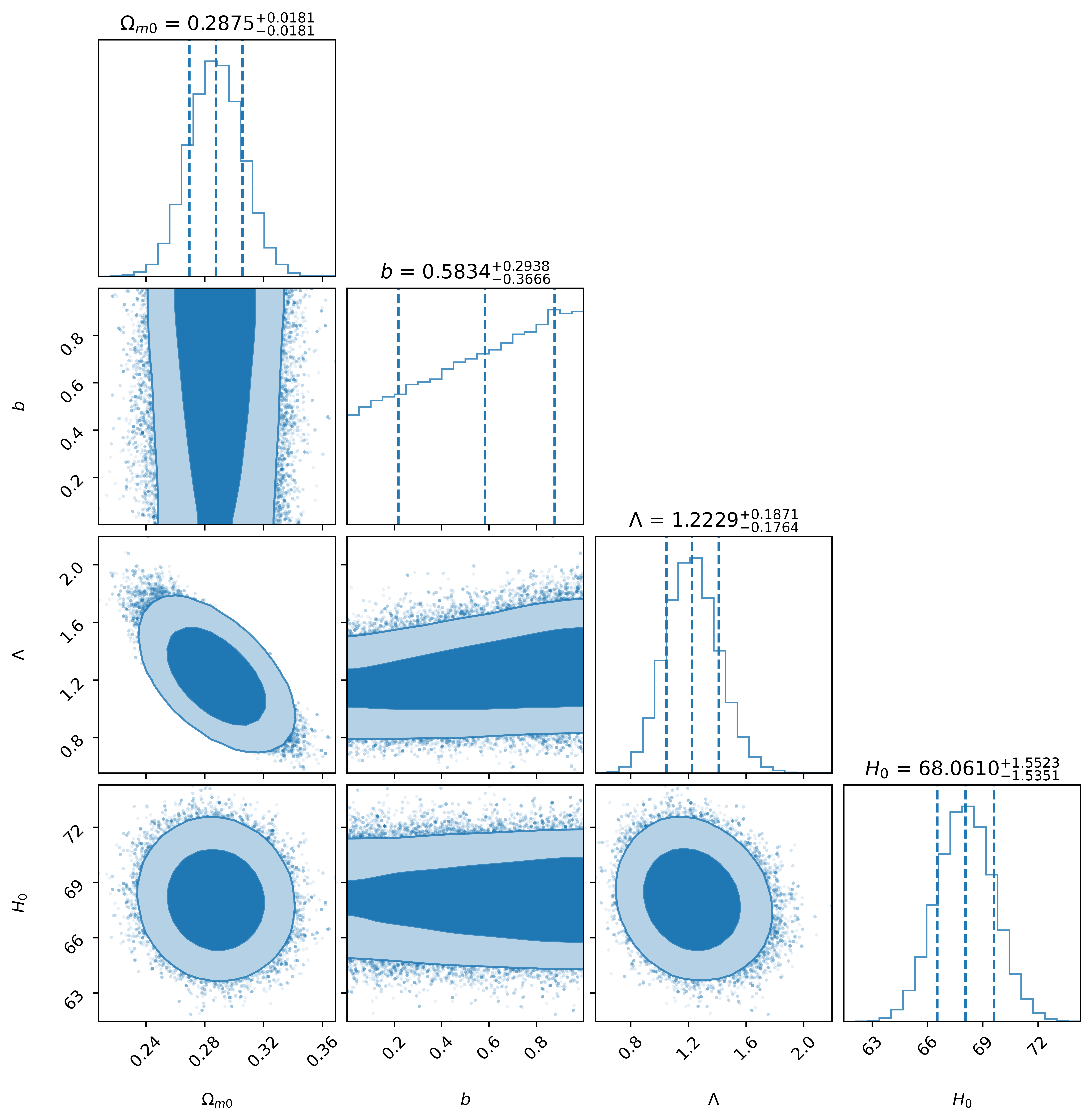}
        \caption{Hu–Sawicki $f(R)$ model.}
        \label{fig:corner_HuSawicki}
    \end{subfigure}
    \hfill
    \begin{subfigure}[t]{0.48\textwidth}
        \centering
        \includegraphics[width=\textwidth]{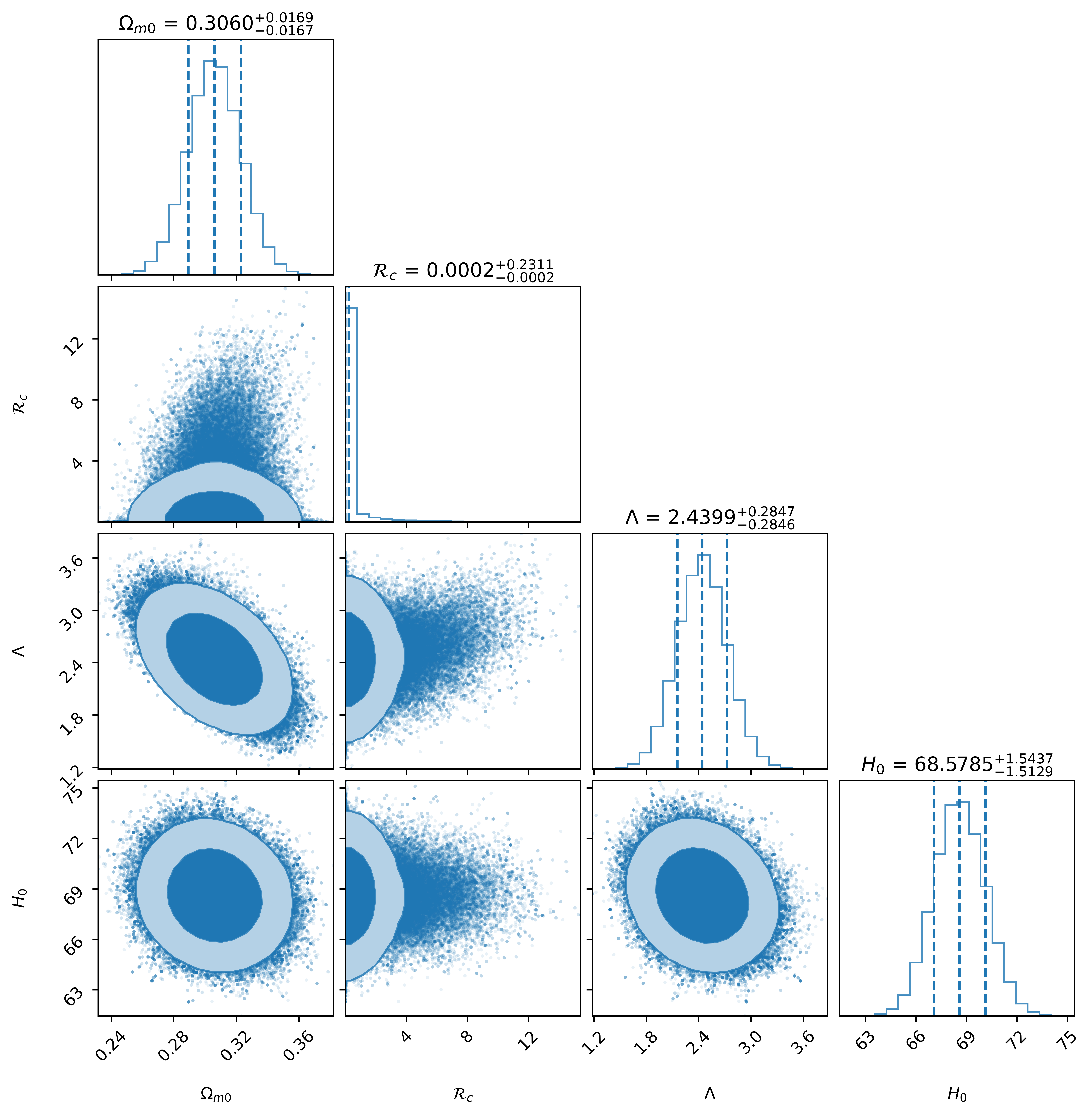}
        \caption{Starobinsky $f(R)$ model.}
        \label{fig:corner_Starobinsky}
    \end{subfigure}

    \caption{Corner plots showing the posterior distributions and covariances  between cosmological parameters  for two $f(R)$ gravity models obtained using Cosmic chronometers $H(z) +$ Unity 3.0 SNe Ia observational datasets. The left panel corresponds to the Hu–Sawicki model having parameters $\Omega_m$, $b$, $\Lambda$, and $H_0$ , while the right panel shows the Starobinsky model having parameters $\Omega_m$, $\mathcal{R}_c$, $\Lambda$, and $H_0$ . Contours represent the 68\% and 95\% confidence levels for each parameter pair.}
    \label{fig:corner_comparison}
\end{figure}

\begin{table}[ht]
\centering
\caption{Comparison of best-fit cosmological parameters for Hu--Sawicki and Starobinsky $f(R)$ gravity models derived from MCMC analysis. Median values with 1$\sigma$ intervals are shown.}
\label{tab:fr_comparison}
\begin{tabular}{lcc}
\toprule
\textbf{Parameter} & \textbf{Hu--Sawicki Model} & \textbf{Starobinsky Model} \\
\midrule
$\Omega_{m}$ & $0.288^{+0.018}_{-0.018}$ & $0.306^{+0.017}_{-0.017}$ \\
$b \text{ or }\mathcal{R}_{c}$ & $b=0.583^{+0.294}_{-0.367}$ & $\mathcal{R}_{c}=2.126\times10^{-4}{}^{+2.31\times10^{-1}}_{-2.12\times10^{-4}}$ \\
$\Lambda$ & $1.229^{+0.187}_{-0.176}$ & $2.440^{+0.285}_{-0.285}$ \\
$H_{0}$ [km s$^{-1}$ Mpc$^{-1}$] & $68.06^{+1.55}_{-1.54}$ & $68.58^{+1.54}_{-1.51}$ \\
\bottomrule
\end{tabular}
\label{tab1}
\end{table}

\begin{figure}[ht]
    \centering
    \begin{subfigure}[t]{0.48\textwidth}
        \centering
        \includegraphics[width=\textwidth]{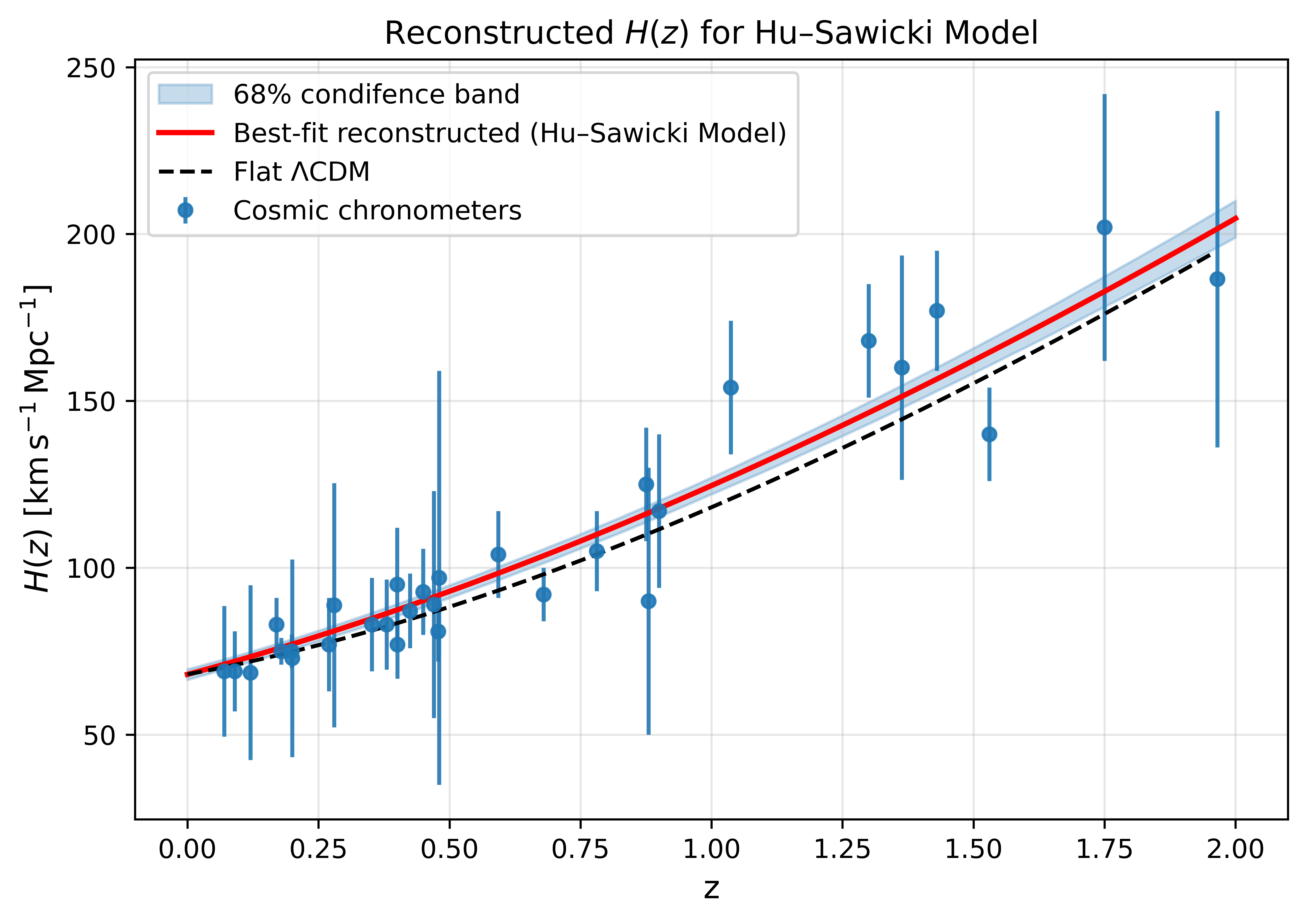}
        \caption{Hu–Sawicki $f(R)$ model.}
        \label{fig:corner_HuSawicki1}
    \end{subfigure}
    \hfill
    \begin{subfigure}[t]{0.48\textwidth}
        \centering
        \includegraphics[width=\textwidth]{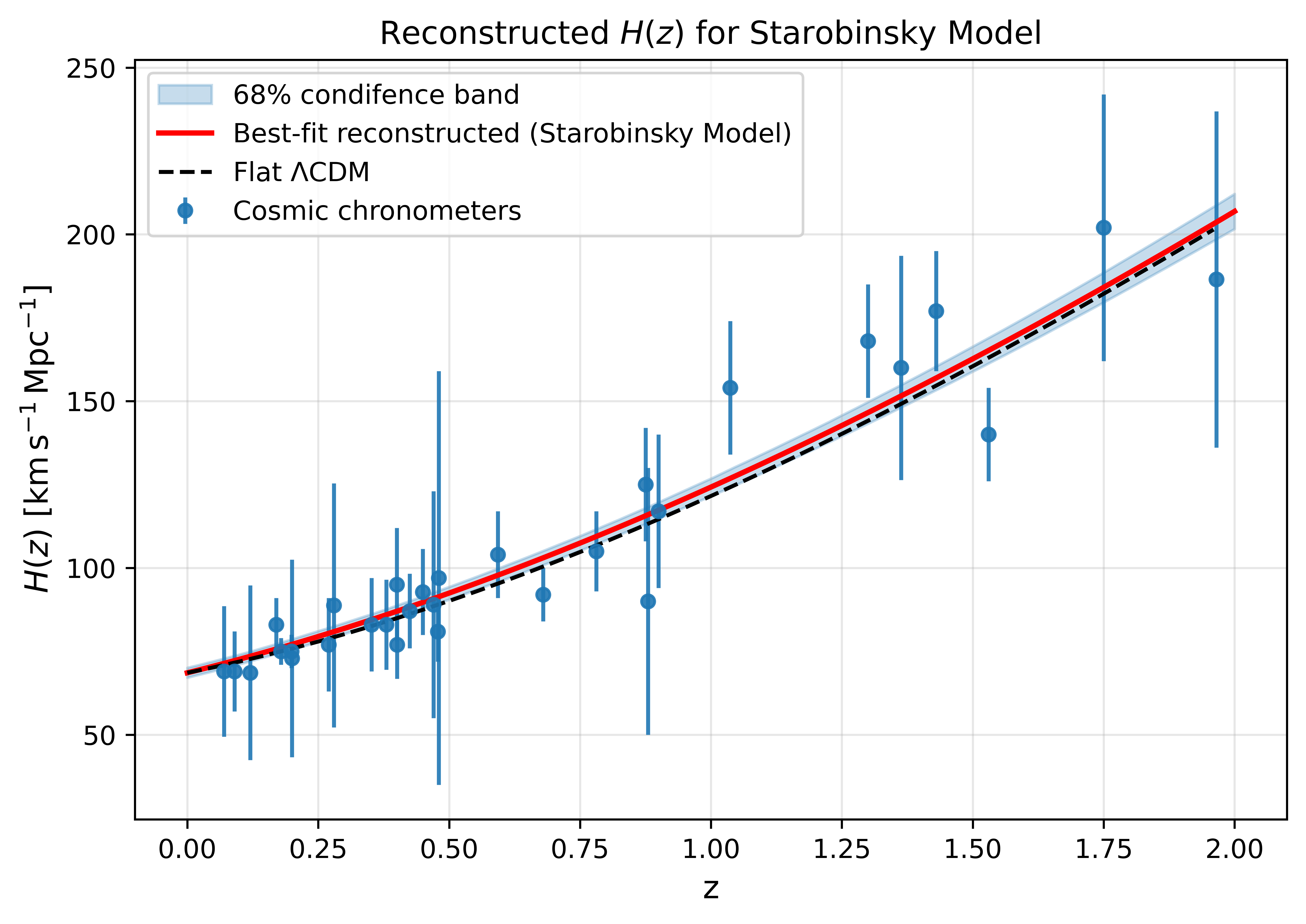}
        \caption{Starobinsky $f(R)$ model.}
        \label{fig:corner_Starobinsky1}
    \end{subfigure}

   \caption{Reconstructed expansion histories $ H(z)$ for two $f(R)$ gravity models compared with the standard $\Lambda$CDM scenario. Panel (a) shows the Hu-Sawicki model, while panel (b) corresponds to the Starobinsky model. The solid blue curve represents the reconstructed median evolution of $H(z)$ with the associated $68\%$ credible region shaded in blue, obtained from the MCMC posterior samples. The dashed black line denotes the best-fit $\Lambda$CDM prediction, and the data points with error bars correspond to cosmic chronometer measurements of $H(z)$ at various redshifts.}

\label{fig:hZ_coparison}
\end{figure}

As result of our MCMC analysis coupled with spectral collocation method, Figure~\ref{fig:corner_comparison} and Table~\ref{tab1} together summarize the statistical outcome  of the joint $H(z)+$Union\,3.0 analysis.  The corner plots in Figure~\ref{fig:corner_comparison} show the posterior probability distributions  and parameter covariances for both benchmark models.  The nearly Gaussian contours in $(\Omega_{m0},H_0)$ demonstrate that the reconstructed  background expansion is tightly constrained and consistent with the Planck~2018  $\Lambda$CDM baseline, while the weak degeneracy between $H_0$ and the model-specific  parameters ($b$ or $R_c$) reflects the limited sensitivity of background probes to higher-order  curvature effects. In both cases, the deviation parameters converge to small but finite values  ($b\simeq0.6$, $R_c/H_0^2\simeq2\times10^{-4}$), showing that current data are compatible  with $\Lambda$CDM yet still allow mild modifications of gravity.  The marginalized posteriors are unimodal and sharply peaked, confirming the numerical  stability of the spectral solver and the absence of multimodal degeneracies that often  appear in perturbative treatments.  Table~\ref{tab1} quantifies these trends, with posterior means of  $(\Omega_{m0},H_0,\Lambda/H_0^2)$ consistent within $1\sigma$ of the concordance values,  while the uncertainties on $b$ and $R_c$ delineate the empirical boundary of  viable $f(R)$ deviations in the late universe.  Together, Figure~\ref{fig:corner_comparison} and Table~\ref{tab1} confirm the statistical  robustness of the spectral Chebyshev reconstruction and its sub-percent recovery of  cosmological parameters.

Figure~\ref{fig:hZ_coparison} provides a direct visual validation of the reconstructed expansion history obtained from the joint $H(z)+{\rm SN}$ analysis. The median spectral $H(z)$ curves (solid blue) trace the observational cosmic-chronometer measurements with residuals below $1\%$ relative to the $\Lambda$CDM prediction (dashed black). The shaded $68\%$ credible bands remain narrow across the full redshift interval $0<z<2$, demonstrating that the Chebyshev collocation method achieves global accuracy and numerical stability without resorting to stepwise numerical integration. The near-complete overlap between the Hu--Sawicki and Starobinsky reconstructions indicates that their background evolutions are observationally degenerate within current precision; any distinction between them must therefore arise from perturbation-level observables such as the growth rate $f\sigma_8(z)$ or weak-lensing convergence. 


Interpreting the reconstructed $\Lambda_{\rm eff}$ as a geometric contribution to cosmic acceleration, we find $\Lambda_{\rm eff}/H_0^2 \simeq 1.23 \pm 0.18$ for the Hu--Sawicki model and $2.44 \pm 0.29$ for the Starobinsky model. The value for the Hu--Sawicki case is clearly lower than the $\Lambda$CDM reference ($\Lambda/H_0^2 \approx 2.21$), which suggests that in this model, part of the observed acceleration arises from the curvature term in $f(R)$ rather than from a large cosmological constant. In other words, the geometry itself contributes to the acceleration, allowing the model to reproduce the same late-time behaviour with a smaller effective vacuum energy. The Starobinsky model, on the other hand, gives a value close to $\Lambda$CDM, indicating that its curvature corrections mainly mimic rather than replace the cosmological constant. Overall, these results show that both models can explain the accelerated expansion without introducing a new dark-energy component, with $\Lambda$CDM appearing as the limiting case when curvature corrections become negligible.

Overall, the combined analysis of the $H(z)$ and Union,3 datasets demonstrates that the spectral Chebyshev approach yields stable and accurate cosmological constraints, giving$(\Omega_{m0},, H_0,, \Lambda_{\rm eff}) \simeq (0.29,, 68,, 1.2$--$2.5,H_0^2)$,in close consistency with the concordance $\Lambda$CDM framework.The small yet finite curvature-induced corrections, characterized by the parameters $b$ and $R_c$, define a physically viable and observationally accessible regime. Forthcoming large-scale surveys such as \emph{Euclid} \cite{euclid_2011}, \emph{LSST} \cite{Ivezic_2019_LSST}, and \emph{SKA} \cite{Braun_2015_SKA} are expected to reach the sensitivity required to probe the growth of cosmic structure and weak-lensing signals, thereby providing a critical test for these subtle departures from General Relativity.

\section{Discussion} \label{sec:discussion_5}
The results of this study demonstrate that the spectral Chebyshev collocation method provides an accurate, stable, and non-perturbative framework for solving the stiff, nonlinear background equations of $f(R)$ cosmology. Applied to the Hu--Sawicki and Starobinsky models, the reconstructed expansion histories $E(z)=H(z)/H_0$ reproduce the observational $H(z)$ data up to $z\simeq2$ with sub-percent residuals relative to $\Lambda$CDM, yielding posterior constraints on $(\Omega_{m0},H_0)$ consistent with Planck and cosmic-chronometer results. The deviation parameters $b$ and $\mathcal{R}_c$ converge to small but finite values, indicating that both models remain viable extensions of $\Lambda$CDM within current observational precision.

It is instructive to contrast this approach with previous solution strategies for the $f(R)$ background equations. The earliest and most direct method has been numerical integration of the modified Friedmann equation using adaptive Runge--Kutta or Bulirsch--Stoer algorithms. Although formally exact, these integrators are highly sensitive to stiffness—particularly near the $\Lambda$CDM limit, where the system involves both slow and rapidly varying modes—often requiring extremely small step sizes or leading to unphysical branches of the solution. To circumvent this issue, perturbative schemes such as the \emph{$b$-expansion} introduced by Basilakos et al.~\cite{Basilakos_2013} approximate the model as a small deformation of the $\Lambda$CDM background, 
\begin{equation}
E^2(z)\simeq E^2_{\Lambda\mathrm{CDM}}(z)+b\,\Delta(z),
\end{equation}
which reproduces the background evolution accurately for $b\ll1$ (or $\mathcal{R}_c^{-1}\!\ll1$ in the Starobinsky case). While computationally efficient, this approach fails in the non-linear regime where deviations from General Relativity become significant—the regime of greatest theoretical interest.

Other semi-analytical formulations have sought to interpolate between low- and high-redshift behaviours. For example, Cardon et al.~\cite{Cardone_2012} proposed an empirical ansatz for the normalized Hubble parameter, Eq.~(2.18) of their work,
\begin{equation}
E(z)=\epsilon(z)E_{\mathrm{CPL}}(z,\Omega_M)+[1-\epsilon(z)]E_\Lambda(z,\Omega_M),
\end{equation}
where $\mathcal{\epsilon}(z)=\sum_{i=1}^{3}e_i(z-z_\Lambda)^i$ ensures a smooth transition between the CPL behaviour at low $z$ and a $\Lambda$CDM--like limit at high $z$.   Although this parametrization achieves sub-percent accuracy in reproducing the numerical solution, it remains purely phenomenological—the coefficients $(e_1,e_2,e_3,z_\Lambda)$ have no direct theoretical correspondence with the underlying model parameters.

Padé or rational-function expansions, sometimes adopted to improve convergence of the $b$-series, suffer from similar drawbacks: local validity, sensitivity to truncation order, and the absence of a direct connection to the governing field equations \cite{Aviles_2014,Sotiriou_2010}. Reconstruction techniques based solely on observational $H(z)$ or supernova data yield model-independent $E(z)$ curves \cite{Sultana_2022,Darshan_2025,Capozziello_2018}, reproduce the observed kinematics of the Universe but they do not connect the recovered expansion history to the fundamental form of the gravitational Lagrangian $f(R)$ or the underlying scalaron dynamics.

In contrast, the Chebyshev collocation method presented here solves the dimensionless master equation directly, preserving the full dependence on $f_{\mathcal{R}}$ and $f_{\mathcal{RR}}$ while ensuring exponential convergence for smooth $E(z)$. By transforming the differential system into a finite set of algebraic constraints for the spectral coefficients, the method eliminates numerical stiffness and attains global accuracy with only eight to ten basis functions. The resulting framework combines the stability of implicit solvers with the precision of spectral approximations, providing a non-perturbative, globally valid solution across the entire redshift interval $0\le z\le z_{\max}$. This property is particularly valuable for cosmological parameter inference, where repeated evaluations within MCMC sampling demand both speed and reliability.

Some limitations remain. The current formulation assumes analytic smoothness of $E(z)$; strongly oscillatory or discontinuous behaviours may require adaptive or piecewise spectral representations. Moreover, this study focuses solely on background dynamics—complete discrimination among modified gravity models demands inclusion of perturbation-level observables such as the linear growth rate, weak-lensing convergence, and CMB anisotropies. Future developments will incorporate first-order scalar and tensor perturbations within the same spectral framework and embed the solver into Boltzmann codes such as \texttt{CLASS} and \texttt{CAMB}, allowing direct computation of matter-power and lensing spectra.

The reconstructed cosmological parameters obtained from the joint $H(z)$ and \textsc{Union\,3.0} supernova datasets yield values fully consistent with the $\Lambda$CDM framework, while retaining small but finite deviations characterized by $b$ in the Hu--Sawicki model and $\mathcal{R}_c$ in the Starobinsky model. These finite departures from the $\Lambda$CDM limit leave a narrow yet meaningful window for exploring extensions of General Relativity that remain observationally viable. The spectral reconstruction therefore not only reproduces the observed expansion history with sub-percent precision but also preserves sensitivity to subtle modifications of gravity at cosmological scales.

In conclusion, the Chebyshev collocation approach establishes a robust and globally convergent numerical framework for $f(R)$ cosmology. Future extensions of this work could include coupling the Chebyshev solver to the linear perturbation equations for structure growth, applying it to other classes of modified gravity such as scalar–tensor or $f(T)$ models, and integrating it into Bayesian inference pipelines.
Ultimately, the framework presented here positions spectral collocation techniques as a  precise, flexible, exact and physically grounded method toward testing gravity beyond the standard $\Lambda$CDM paradigm.

\section*{Acknowledgement}The author is thankful to his undergraduate Numerical Analysis class at St. Stephen’s College, whose discussions helped spark the idea that led to this work.The author wishes to express sincere gratitude to the Principal of St. Stephen’s College and to the Centre for Theoretical Physics, St. Stephen’s College, for their continuous support and for providing the necessary academic infrastructure. The author also gratefully acknowledges Prof. Shobhit Mahajan and Prof. Deepak Jain for their valuable guidance.

\bibliographystyle{JHEP}
\bibliography{reference}

\appendix
\section*{Appendix: Numerical Solution of the Dimensionless Master Equation Using Chebyshev Collocation}
\addcontentsline{toc}{section}{Appendix: Numerical Solution of the Dimensionless Master Equation Using Chebyshev Collocation}
\label{app:cheb_solver}

This appendix describes in detail the numerical procedure used to solve the dimensionless master equation
\[
\mathcal{F}\big(E(z),E'(z),E''(z); z,\Omega_{m0},\theta\big) = 0,
\]
for a fixed set of model parameters (for example, the Hu--Sawicki case with \(b=0.583\) and \(\tilde{\Lambda}=1.229\)). 
The procedure is constructive: we first represent the normalized Hubble function \(E(z)\) through a Chebyshev expansion, then define collocation nodes and corresponding differentiation operators, and finally assemble the discrete algebraic system obtained by enforcing the master equation at the collocation points.

Throughout this appendix, we illustrate the steps explicitly for 
\[
N=8, \qquad z_{\max}=30,
\]
which are representative of the configurations used in the numerical experiments presented in the main text.

\subsection*{Step 1: Chebyshev Expansion of \(E(z)\)}

We approximate the normalized Hubble function \(E(z)\) by a truncated Chebyshev series:
\begin{equation}
E(z) \simeq \sum_{k=0}^{N} a_k\,T_k(x(z)),
\label{eq:E_cheb_app}
\end{equation}
where \(T_k(x)\) is the \(k\)-th Chebyshev polynomial of the first kind, and the variable \(x \in [-1,1]\) is linearly related to redshift \(z\in[0,z_{\max}]\) through
\begin{equation}
x = \frac{2z}{z_{\max}} - 1, 
\qquad \Longleftrightarrow \qquad 
z = \frac{z_{\max}}{2}(x+1).
\label{eq:xz_map_app}
\end{equation}
For \(N=8\) and \(z_{\max}=30\), the expansion explicitly reads
\[
E(z) \simeq a_0 T_0(x) + a_1 T_1(x) + \dots + a_8 T_8(x),
\]
with recursive relation

\[
T_0(x)=1,\qquad T_1(x)=x,
\]
\[
T_{n+1}(x)=2x\,T_n(x)-T_{n-1}(x),\qquad n\ge 1.
\]

\subsection*{Step 2: Collocation Nodes (Chebyshev--Gauss--Lobatto)}

The collocation nodes are chosen as the Chebyshev--Gauss--Lobatto points:
\begin{equation}
x_j = \cos\!\left(\frac{\pi j}{N}\right), \qquad j = 0,1,\dots,N,
\label{eq:CGL_app}
\end{equation}
which are then mapped to redshift space using Eq.~\eqref{eq:xz_map_app}:
\begin{equation}
z_j = \frac{z_{\max}}{2}(x_j + 1).
\label{eq:z_nodes_app}
\end{equation}
For \(N=8\) and \(z_{\max}=30\), the collocation nodes are:

\[
\begin{tabular}{c c c}
\toprule
$j$ & $x_j$ & $z_j$ \\
\midrule
0 & 1.00000000 & 30.000000 \\
1 & 0.92387953 & 28.858193 \\
2 & 0.70710678 & 25.606602 \\
3 & 0.38268343 & 20.740251 \\
4 & 0.00000000 & 15.000000 \\
5 & -0.38268343 & 9.259749 \\
6 & -0.70710678 & 4.393398 \\
7 & -0.92387953 & 1.141807 \\
8 & -1.00000000 & 0.000000 \\
\bottomrule
\end{tabular}
\]
These nodes include both endpoints of the interval and are denser near the boundaries, which improves the spectral representation of smooth functions.

\subsection*{Step 3: Differentiation Matrices and Derivatives of \(E(z)\)}

In this step, we derive the differentiation matrices that allow us to compute the first and second derivatives of the normalized Hubble function \(E(z)\) directly at the Chebyshev--Gauss--Lobatto nodes introduced in Step~2.  
These matrices form the backbone of the Chebyshev collocation method and replace analytical differentiation by accurate matrix operations.

\bigskip

\noindent
\textbf{Concept:}  
Suppose we have a smooth function \(f(x)\) whose values are known at the \(N+1\) collocation nodes \(\{x_j\}\).  
We can construct its interpolating polynomial \(p(x)\) of degree \(N\) that passes exactly through all those points:
\[
p(x) = \sum_{j=0}^{N} f(x_j)\,L_j(x),
\]
where \(L_j(x)\) are the \emph{Lagrange basis polynomials} satisfying the cardinal property
\[
L_j(x_i) = \delta_{ij}.
\]
The derivative of this interpolating polynomial is then
\[
p'(x) = \sum_{j=0}^{N} f(x_j)\,L_j'(x).
\]
If we evaluate this derivative at each node \(x_i\), we obtain
\[
p'(x_i) = \sum_{j=0}^{N} L_j'(x_i)\,f(x_j).
\]
Thus, the derivative of the function at the collocation nodes can be written in matrix form as
\[
\mathbf{f}' = D_x\,\mathbf{f}, \qquad 
(D_x)_{ij} = L_j'(x_i),
\]
where \(\mathbf{f} = [f(x_0),f(x_1),\dots,f(x_N)]^{\mathsf{T}}\) and \(D_x\) is called the \emph{differentiation matrix}.  
Each element \((D_x)_{ij}\) measures how much the derivative at node \(x_i\) depends on the value of the function at node \(x_j\).

\bigskip

\noindent
\textbf{Derivation for Chebyshev--Gauss--Lobatto nodes.}  
The Lagrange basis functions \(L_j(x)\) can be written explicitly as
\[
L_j(x) = \prod_{\substack{m=0 \\ m\ne j}}^{N} \frac{x - x_m}{x_j - x_m}.
\]
Directly differentiating this expression is tedious, but for the special case of Chebyshev--Gauss--Lobatto nodes
\[
x_j = \cos\!\left(\frac{\pi j}{N}\right), \qquad j=0,1,\dots,N,
\]
the result simplifies beautifully because of the symmetry and orthogonality properties of the Chebyshev polynomials.  
The resulting closed-form expression for the differentiation matrix entries is:
\[
(D_x)_{jk} =
\begin{cases}
\dfrac{c_j}{c_k}\dfrac{(-1)^{j+k}}{x_j-x_k}, & j\ne k, \\[8pt]
-\displaystyle\sum_{\substack{k=0 \\ k\ne j}}^{N}(D_x)_{jk}, & j=k,
\end{cases}
\]

The scaling factors \(c_j\) account for the unequal spacing of the Chebyshev--Gauss--Lobatto nodes 
\[
x_j = \cos\!\left(\frac{\pi j}{N}\right), \qquad j=0,1,\dots,N.
\]
At the endpoints \(x_0=1\) and \(x_N=-1\), the nodes contribute with half the effective weight of the interior points when differentiating the Lagrange basis functions.  
To preserve the symmetry and accuracy of the derivative operator, these endpoint weights are incorporated as
\[
c_0 = c_N = 2, \qquad c_j = 1 \quad (1 \le j \le N-1).
\]
Thus, \(c_j\) serve to normalize the differentiation matrix and ensure it remains exact for all polynomials up to degree \(N\).

\bigskip

\noindent

The matrix \(D_x\) acts as a discrete version of the derivative operator.  
If the function values are arranged in a column vector \(\mathbf{f}\), then \(D_x\,\mathbf{f}\) gives the derivative of the interpolating polynomial \(p'(x)\) evaluated exactly at all collocation nodes.  
This method is \emph{exact} for any polynomial of degree up to \(N\) and spectrally accurate for smooth functions.  
The off-diagonal elements represent coupling between nodes (since polynomial interpolation is global), while the diagonal elements are chosen so that the derivative of a constant function is zero, ensuring
\[
D_x\,\mathbf{1} = \mathbf{0}.
\]

\bigskip

\noindent
\textbf{Transformation to redshift space.}  
Our physical problem is formulated in terms of the redshift variable \(z\in[0,z_{\max}]\), which is linearly related to \(x\in[-1,1]\) via
\[
x = \frac{2z}{z_{\max}} - 1 \qquad \Longleftrightarrow \qquad z = \frac{z_{\max}}{2}(x+1).
\]
Hence,
\[
\frac{dx}{dz} = \frac{2}{z_{\max}}.
\]
Applying the chain rule gives the differentiation operators in \(z\)-space:
\begin{align}
D^{(1)}_z &= \frac{2}{z_{\max}} D_x, &
D^{(2)}_z &= \left(\frac{2}{z_{\max}}\right)^2 D_x^2.
\end{align}
These matrices allow us to evaluate the first and second derivatives of \(E(z)\) directly as matrix products.

\bigskip

\noindent
\textbf{Matrix representation of the Chebyshev expansion and derivatives.}  

The nodal approximation of the normalized Hubble function can be expressed as
\[
\mathbf{E} =
\begin{bmatrix}
E(z_0) \\[3pt]
E(z_1) \\[3pt]
\vdots \\[3pt]
E(z_8)
\end{bmatrix}
=
\underbrace{
\begin{bmatrix}
T_0(x_0) & T_1(x_0) & \cdots & T_8(x_0) \\[4pt]
T_0(x_1) & T_1(x_1) & \cdots & T_8(x_1) \\[4pt]
\vdots   & \vdots   & \ddots & \vdots   \\[4pt]
T_0(x_8) & T_1(x_8) & \cdots & T_8(x_8)
\end{bmatrix}_{9\times 9}
}_{T_{jk}=T_k(x_j)}
\begin{bmatrix}
a_0 \\[3pt]
a_1 \\[3pt]
\vdots \\[3pt]
a_8
\end{bmatrix}_{9\times 1},
\]
where each \(T_{jk}=T_k(x_j)=\cos(k\arccos x_j)\) is evaluated at the Chebyshev--Gauss--Lobatto nodes
\[
x_j = \cos\!\left(\frac{\pi j}{8}\right), \qquad \text{, where }
z_j = \frac{z_{\max}}{2}(x_j+1), \qquad z_{\max}=30.
\]

Thus, the discrete nodal relation is
\[
E(z_j) = \sum_{k=0}^{8} a_k\,T_k(x_j),
\qquad \text{for } j=0,1,\dots,8,
\]
or equivalently, in compact matrix form,
\[
\boxed{\;\mathbf{E}(z) = T(x)\,\mathbf{a},\quad T_{jk}=T_k(x_j)\;}.
\]


Then the derivatives of \(E(z)\) at the collocation nodes are obtained as
\[
\mathbf{E}' = D^{(1)}_z\,T\,\mathbf{a}, \qquad
\mathbf{E}'' = D^{(2)}_z\,T\,\mathbf{a}.
\]

\bigskip
\noindent 
For these parameter values, the compute the $D^{(1)}_z$ and $D^{(2)}_z$   matrices.

\noindent
 
In the numerical implementation, these matrices are used to compute the derivatives of \(E(z)\) at the collocation nodes:
\[
\mathbf{E}' = D^{(1)}_z\,\mathbf{E}, \qquad 
\mathbf{E}'' = D^{(2)}_z\,\mathbf{E}.
\]
This allows the continuous differential equation to be replaced by a discrete algebraic system that can be solved efficiently.  
Because the Chebyshev method achieves exponential convergence for smooth functions, very high accuracy is obtained even for small values of \(N\).

\bigskip

\noindent
\textbf{{Step 4. Express auxiliary quantities at nodes.} \\}

\noindent
In the next step, we compute all auxiliary quantities required for evaluating the nonlinear master equation at the collocation nodes.We evaluate the dimensionless Ricci curvature $\mathcal{R}(z)$ and its redshift derivative $\mathcal{R}'(z)$ using
\[
\mathcal{R}(z) = 6\Big(2E^2 - (1+z)E\,E'\Big),
\qquad
\mathcal{R}'(z) = 6\Big[4E\,E' - (1+z)(E')^2 - (1+z)E\,E''\Big].
\]
These expressions are computed componentwise at each collocation node $z_j$ using $E_j,E'_j,E''_j$ to obtain $\mathcal{R}_j$ and $\mathcal{R}'_j$.

At each node we must also evaluate $\tilde f(\mathcal{R})$, $\tilde f_{\mathcal{R}}(\mathcal{R})$, and $\tilde f_{\mathcal{RR}}(\mathcal{R})$ for the chosen $f(R)$ model.  
For instance, for the Hu--Sawicki family, the nondimensionalized model is written as
\[
\tilde f(\mathcal{R}) = \mathcal{R} - 2\tilde\Lambda \!\left(1 - \frac{1}{1 + (\mathcal{R}/b\tilde\Lambda)^n}\right),
\]
where $\tilde\Lambda = \Lambda/H_0^2$ is the dimensionless cosmological constant, 
$b$ is a dimensionless deviation parameter, and $n$ is a positive integer. 
For simplicity of illustration, taking $n=1$, the corresponding derivatives are
\[
\tilde f_{\mathcal{R}} = 1 - \frac{2\tilde\Lambda}{b\tilde\Lambda + \mathcal{R}},
\qquad
\tilde f_{\mathcal{RR}} = \frac{2\tilde\Lambda}{(b\tilde\Lambda + \mathcal{R})^2}.
\]
These quantities are evaluated elementwise at the nodal curvature values $\mathcal{R}_j$ and collectively form the vectors for any given value of model parameter, like $(b,\tilde{\Lambda})$ where each entry corresponds to a collocation node $z_j$.

Substituting these into the dimensionless master equation 
\[
\mathcal{F}\big(E,E',E'';\,z,\Omega_{m0},b,\tilde\Lambda\big) = 0,
\]
and enforcing it at every collocation node $z_j$, yields the nonlinear algebraic system
\[
\mathbf{F}(\mathbf{a}) = 
\mathcal{F}\big(T\mathbf{a},D^{(1)}_z T\mathbf{a},D^{(2)}_z T\mathbf{a};\,\mathbf{z},\Omega_{m0},b,\tilde\Lambda\big) = \mathbf{0}.
\]
\noindent
The resulting algebraic system, composed of \(N+1\) nonlinear equations in the \(N+1\) unknown Chebyshev coefficients \(\{a_k\}\), 
is solved iteratively for any chosen parameter pair \((b,\tilde\Lambda)\), 
thereby yielding the approximate spectral solution \(E(z)\) corresponding to the selected \(f(R)\) model. 

The system can be expressed compactly as
\[
\mathbf{F}(\mathbf{a}) = \mathbf{0},
\]
where each component of \(\mathbf{F}\) represents the residual of the dimensionless master equation evaluated at a collocation node.

\bigskip

\noindent\textbf{Step 5: Solving the system of algebric equations.}\\

\noindent
The nonlinear system is solved using a robust root-finding algorithm. 
In this work, we employ the \emph{Powell--hybrid method} 
implemented in \texttt{scipy.optimize.root} with \texttt{method='hybr'}, 
which combines the advantages of the Newton–Raphson and quasi-Newton approaches to achieve rapid convergence.
Convergence is declared when both the infinity norm of the residual and the relative update in the coefficient vector satisfy
\[
\|\mathbf{F}\|_{\infty} < 10^{-10}, 
\qquad 
\frac{\|\Delta\mathbf{a}\|}{\|\mathbf{a}\|} < 10^{-8}.
\]

\noindent
Before solving for the coefficients, the function \(E(z)\) is explicitly normalized at the present epoch by enforcing the boundary condition
\[
E(0) = 1.
\]
This condition replaces the master equation at the collocation node corresponding to \(z=0\) in the residual vector, thereby fixing the overall scale of the solution. 
Consequently, the recovered coefficients \(\{a_k\}\) yield a normalized Hubble function \(E(z)=H(z)/H_0\), 
and no additional scaling is required during evaluation.

\bigskip

\noindent\textbf{Step 6: Numerical solution and explicit form of \(E(z)\).}\\

\noindent Using the Chebyshev collocation setup described above and solving the nonlinear algebraic system for the parameter choice
\[
\Omega_{m0}=0.288,\qquad b=0.583,\qquad \tilde\Lambda=1.229,\qquad z_{\max}=30,\qquad N=8,
\]
we obtain the Chebyshev coefficients (rounded to nine significant figures)
\[
\mathbf{a} =
\begin{bmatrix}
a_0\\ a_1\\ a_2\\ a_3\\ a_4\\ a_5\\ a_6\\ a_7\\ a_8
\end{bmatrix}
=
\begin{bmatrix}
40.4098511\\
46.5491027\\
6.20348851\\
-0.660419935\\
0.174564824\\
-0.0620043498\\
0.0245098272\\
-0.0106977736\\
0.00356633984
\end{bmatrix}.
\]

The normalized Hubble function is therefore represented by the Chebyshev series
\[
E(z) \simeq \sum_{k=0}^{8} a_k\,T_k\!\big(x(z)\big),
\qquad x(z)=\frac{2z}{z_{\max}}-1=\frac{2z}{30}-1,
\]
which for the above coefficients becomes explicitly
\[
\boxed{%
\begin{aligned}
E(z) \simeq\;&
40.4098511\,T_0\!\big(x(z)\big)
+46.5491027\,T_1\!\big(x(z)\big)
+6.20348851\,T_2\!\big(x(z)\big)\\[4pt]
&-0.660419935\,T_3\!\big(x(z)\big)
+0.174564824\,T_4\!\big(x(z)\big)
-0.0620043498\,T_5\!\big(x(z)\big)\\[4pt]
&+0.0245098272\,T_6\!\big(x(z)\big)
-0.0106977736\,T_7\!\big(x(z)\big)
+0.00356633984\,T_8\!\big(x(z)\big),
\end{aligned}
}
\]

Here \(T_k(x)\) are the Chebyshev polynomials of the first kind . These coefficients (and the resulting \(E(z)\)) are the solution for the chosen parameter set \((\Omega_{m0},b,\tilde\Lambda)\). For different parameter choices the same numerical procedure (Steps 1–5) is used to produce a different coefficient vector \(\mathbf{a}\) and hence a different spectral approximation for \(E(z)\).

\end{document}